\documentclass[journal]{IEEEtran}

\usepackage{cite}
\usepackage{graphicx}
\usepackage[cmex10]{amsmath}
\usepackage{amssymb}
\usepackage{flushend}

\begin{document}
%
\title{Assigning UPDRS Scores in the Leg Agility Task of Parkinsonians: Can It Be Done through BSN-based Kinematic Variables?}
%
%
%

\author{Matteo Giuberti,~\IEEEmembership{Student Member,~IEEE,} Gianluigi Ferrari,~\IEEEmembership{Senior Member,~IEEE,} Laura Contin, Veronica Cimolin, Corrado Azzaro, Giovanni Albani, and Alessandro Mauro
\thanks{M. Giuberti is now with Xsens Technologies B.V., 7500 AN Enschede, The Netherlands. E-mail: matteo.giuberti@xsens.com. He was
with CNIT Research Unit of Parma and the Department of Information Engineering of the University of Parma, I-43124 Italy, when this work was performed.}
\thanks{G. Ferrari is with CNIT Research Unit of Parma and the Department of Information Engineering of the University of Parma, I-43124, Italy. E-mail: gianluigi.ferrari@unipr.it.}
\thanks{L. Contin is with Research \& Prototyping, Telecom Italia, Turin, Italy. E-mail: laura.contin@telecomitalia.it}
\thanks{V. Cimolin is with the Department of Electronics, Information, and Bioengineering, Polytechnic of Milan, Italy. E-mail: veronica.cimolin@biomed.polimi.it}
\thanks{C. Azzaro and G. Albani are with the Division of Neurology and Neurorehabilitation, Istituto Auxologico Italiano IRCCS, Piancavallo (VB), Italy. E-mail: \{c.azzaro,g.albani\}@auxologico.it}
\thanks{A. Mauro is with the Division of Neurology and Neurorehabilitation, Istituto Auxologico Italiano IRCCS, Piancavallo (VB), Italy and with the Department of Neurosciences, University of Turin, Italy. E-mail: mauro@auxologico.it}
}

%
%

\markboth{IEEE Internet of Things Journal, Special Issue on ``Internet of Things for Smart and Connected Health,'' June 2014}%
{Giuberti \MakeLowercase{\textit{et al.}}: Assigning UPDRS Scores in the Leg Agility Task...}
%



\maketitle

\begin{abstract}
In this paper, by characterizing the Leg Agility (LA) task, which contributes to the evaluation of the degree of severity of the Parkinson's Disease (PD), through kinematic variables (including the angular amplitude and speed of thighs' motion), we investigate the link between these variables and Unified Parkinson's Disease Rating Scale (UPDRS) scores. Our investigation relies on the use of a few body-worn wireless inertial nodes and represents a first step in the design of a portable system, amenable to be integrated in Internet of Things (IoT) scenarios, for automatic detection of the degree of severity (in terms of UPDRS score) of PD. The experimental investigation is carried out considering 24 PD patients.
\end{abstract}


%
\IEEEpeerreviewmaketitle

\section{Introduction}
\subsection{Motivation}
\IEEEPARstart{P}{arkinson}'s Disease (PD) is the second most common neurodegenerative disorder after Alzheimer's disease. According to the Global Declaration for Parkinson's Disease, 6.3 million people suffer from PD worldwide~\cite{epda}. The prevalence of PD is about 0.3\% of the whole population in industrialized countries, rising up to 1\% over the age of 65 and to 4\% over 80. The clinical picture of PD is characterized by a progressive deterioration of the motor performance, with the occurrence of slowness (bradykinesia) and poverty of voluntary movements, expressionless face, ``resting'' tremor, stooped posture, festinating gait, axial instability. Asymmetry of motor symptoms is a typical feature of PD. Although the symptoms can be improved by dopaminergic drugs, such as L-dopa, over time its effectiveness worsens and motor fluctuations may occur as well as dyskynesias and involuntary movements. Furthermore, variations in the severity of these symptoms are observed during dosing intervals. 

The clinical picture assessed during an outpatient check up in the medical office poorly represents the real (actual) clinical status, especially in fluctuating patients. Indeed, repeated daily assessments of motor symptoms would be required and this is usually done by asking the patient to annotate the number of hours of OFF (i.e., when drugs are not effective) and ON condition (i.e., when they are effective), but this is not fully reliable due to perceptual bias. For this reason, in recent years a number of studies on automatic systems to evaluate motor fluctuations of PD patients have been developed~\cite{ChPaBuReMcShTaWeBo11}. The most common approach is leveraging sensing technology to automatically evaluate the performance of specific motor tasks, such as ``sit-to-stand''~\cite{NuLaVaHi86,GiMaBeMa07}, gait analysis~\cite{SaRuViDeBlBuAm04,SaZaHoCaNuAm09}, and tremor~\cite{SaRuWiBuViAm07}. The basic idea is to develop a system able to get an evaluation of the motor status of a patient as close as possible to the evaluation of neurologists when they apply semi-quantitative evaluation scales, such as the Unified Parkinson's Disease Rating Scale (UPDRS)~\cite{FaEl87}.

\subsection{The Leg Agility Task} 
Although several works have appeared focusing on the evaluation of the performance of specific motor tasks, such as ``sit-to-stand''~\cite{NuLaVaHi86,GiMaBeMa07}, gait analysis~\cite{SaRuViDeBlBuAm04,SaZaHoCaNuAm09}, and tremors~\cite{SaRuWiBuViAm07}, to the best of our knowledge, limited attention has been devoted, in the literature, to the evaluation of the Leg Agility (LA) task~\cite{PaLoHuHuGrStAkDyWeBo09,DaTrMuAlOhDeHo11,HeFiRiWhWaGuGiMe12,GiFeCoCiCaGaAzAlMa13,GiFeCoCiAzAlMa14}.

\subsubsection{Task Description} \label{sec:LA_description}
The LA task aims at evaluating the severity of motion impairments of a PD patient, with specific focus on the lower limbs. In this exercise, the patient is asked to sit on a chair provided with rigid backrest and armrests. The patient must place both his/her feet on the floor in a comfortable position. The exercise consists in alternately raising up and stomping the feet on the ground, as high and as fast as possible. Ten repetitions per leg must be performed while sitting on the chair in order to test each leg separately. The examiner should first train the patient, showing him/her the correct execution of the exercise, stopping as soon as the patient starts. The significant parameters that have to be measured, independently for each leg, are the speed, the regularity, and the amplitude of the movement. Moreover, differences can be observed between the movements performed with the different legs. For this reason, in the following we will distinguish between Right LA (RLA) and Left LA (LLA) tasks.

\subsubsection{UPDRS Evaluation}
According to the guidelines of the Movement Disorder Society (MDS), the LA task must be evaluated observing the following parameters: amplitude, slowing, hesitations, interruptions, and freezing. In particular, in Table~\ref{tab:mappingUPDRS}, an attempt at mapping these parameters with an UDPRS evaluation is presented.
\begin{table*}
\caption{UPDRS mapping.}
\label{tab:mappingUPDRS}
\normalsize
\centering
\begin{tabular}{|c||c|c|c|c|c|c|}
\hline
{\bf UPDRS} & {\bf Amplitude} & {\bf Slowing} & {\bf Hesitations} & {\bf Interruptions} & {\bf Freezing}\\
\hline
\hline
0 & nearly constant & no & 0 & 0 & 0 \\
\hline
1 & decrements near the end & slight & $\geq1$ & 1,2 & 0\\
\hline
2 & decrements midway & mild & - & 3,4,5 & 0\\
\hline
3 & decrements after first tap & moderate & - & $\geq6$ & $\geq1$\\
\hline
4 & always minimal or null & severe & - & always & - \\
\hline
\end{tabular}
\end{table*}
To this end, note that UPDRS scores are integer values that range from 0 (no problems) to 4 (worst conditions).
While the first feature (i.e., amplitude) directly corresponds to a physical measure, the quantitative evaluation of the other ones typically relies on the experience of neurologists. Therefore, inter-neurologist score variations cannot be a priori excluded.

\subsection{Paper Contribution}
\label{subsec:paper_cont}
In this work, we focus on the characterization of the LA task in PD patients, devising an approach for quantitative evaluation of relevant kinematic features representative of the UPDRS score of a
patient. This work extends the preliminary results presented in~\cite{GiFeCoCiCaGaAzAlMa13,GiFeCoCiAzAlMa14}, including also a novel frequency domain
analysis, which allows to identify more accurately relevant kinematic features representative
of the UPDRS level. After showing, with direct comparison with an optoelectronic system,  that the LA task can be effectively characterized by analyzing the inclination and angular velocity of the thighs, we characterize the kinematic variables associated with the thighs' motion. We first present a ``single subject'' experimental characterization the LA task, comparing directly a healthy patient with a PD patient, highlighting similarities and differences. Then, we perform a ``large-scale'' (considering 24 PD patients) experimental analysis, identifying the most significant  kinematic features associated with the LA task characterization, by mapping them with the UPDRS scores attributed by expert neurologists. 
The encouraging experimental results, suggesting that the UPDRS score might be concisely interpreted, e.g., as decreasing function of the ``power'' of a movement, motivate the design and implementation of an automatic UPDRS detection system based on the use of a few body-worn inertial sensors. Moreover, the use of wireless inertial sensors makes such a portable system easily integrable in Internet of Things (IoT) scenarios, allowing directly remote monitoring and data sharing.

\subsection{Paper Structure}
\label{subsec:paper_struc}
This paper is structured as follows.
Section~\ref{sec:exp_setup} describes the experimental set-up, detailing the used hardware and
the considered subjects (both Parkinsonians and healthy).
In Section~\ref{sec:la_chara}, the LA task is characterized by, first, showing
the equivalence between heel elevation (as measured with an optoelectronic system) and
thigh inclination (as measured with the wireless inertial sensor-based system) and, then,
by extracting relevant kinematic features, in both time and frequency domains.
In Section~\ref{sec:results}, the obtained experimental results are presented and commented. 
Finally, Section~\ref{sec:conc} concludes the paper.

\section{Experimental Set-up} \label{sec:exp_setup}

\subsection{Hardware Description} \label{subsec:hw}
The experiments were carried out at the San Giuseppe Hospital, Istituto Auxologico
Italiano, in Piancavallo (Verbania, Italy), at a fully equipped last generation motion analysis laboratory. In particular, the kinematic analysis was carried out, in a comparative way, considering (i)~an optoelectronic system and (ii)~a wireless Body Sensor Network (BSN)-based system, based on a few nodes (equipped with inertial and magnetic sensors) placed over the body.

The optoelectronic system (Vicon, Oxford, UK) performs a real-time processing of images from 6 fixed infrared cameras (with sampling rate equal to 100~Hz) to extract the reflectance of passive markers (with a diameter of 15~mm) which are positioned on specific anatomical landmarks of the subject. Prior to testing, the system was calibrated to assure accuracy and to allow the computation of each marker's 3D coordinates. The average error on the computation of the difference between measured and actual distances of two markers fixed at the edges of a rigid bar was within 0.21~mm (with standard deviation equal to 0.1~mm). The calibrated volume for this application had: length equal to 3.5~m ($x$ axis of the laboratory reference system); height equal to 2~m ($y$ axis of the laboratory reference system); and width equal to 2~m ($z$ axis of the laboratory reference system).

The BSN is formed by Shimmer (Sensing Health with Intelligence, Modularity, Mobility, and Experimental Reusability) nodes~\cite{BuGrMcOSKuAyStCi10}.
A Shimmer node is a small and low-power wireless sensing platform that can capture and communicate a wide range of sensed data in real time.
The main module is a compact wearable device (size: 53mm x 32mm x 25mm; weight: 22g) equipped with: a TI MSP430 microcontroller; a Bluetooth radio (Roving Networks RN-42) and an IEEE 802.15.4 compliant radio (TI CC2420); an integrated 2 GB microSD card slot; a 450mAh rechargeable Li-ion battery; and a triaxial accelerometer (Freescale MMA7361). Moreover, the device is designed so that different external sensing modules can be easily connected. In particular, the 9DoF Kinematic Sensor expansion module, which is supplied with a triaxial gyroscope (InvenSense 500 series) and a triaxial magnetometer (Honeywell HMC5843), has been used.

\subsection{Subjects} \label{subsec:subjects}
\subsubsection{Single-subject Analysis} \label{subsubsec:single_sub}
We first evaluated the LA tasks performed by two individuals: one healthy subject (subject A) and a PD patient (subject B). Subject B has a disease duration of 4 years and does not present motor fluctuations. His Hoehn \& Yahr score was 2 and the UPDRS score for LA was 1 bilaterally. The score of 1 was assigned because of the presence of one hesitation of the movement and a slight slowing during the limb motion. In Table~\ref{tab:subjects}, we summarize the data of the considered subjects, indicating also the performed exercises.
\begin{table*}[t]
\caption{Considered subjects for a direct (one-to-one) comparison between PD and healthy subjects.}
\label{tab:subjects}
\normalsize
\centering
\begin{tabular}{|c|c|c|c|c|c|c|}
\hline
{\bf Subject} & {\bf Sex} & {\bf Age} & {\bf Weight} & {\bf Height} & {\bf UPDRS score for LA} & {\bf Exercises}\\
\hline
\hline
A (healthy) & female & 40 & 56~Kg & 171~cm & 0 & 1xRLA, 1xLLA\\
\hline
B (PD) & male & 42 & 85~Kg & 180~cm & 1 (bilaterally) & 1xRLA, 1xLLA\\
\hline
\end{tabular}
\end{table*}

\subsubsection{Large-scale Analysis}
\begin{figure}[t]
\centering
\includegraphics[height=0.19\textheight]{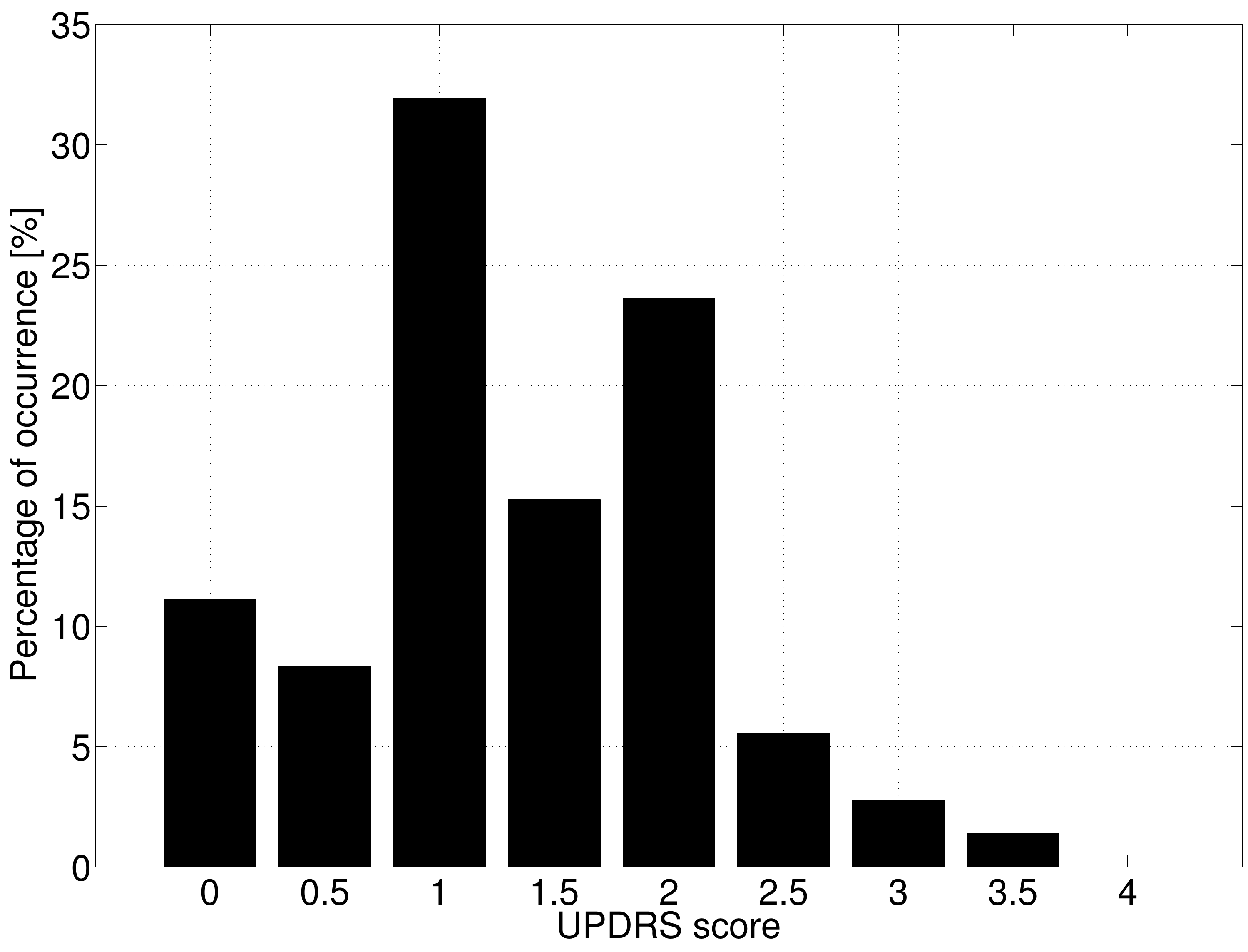}
\caption{Distribution of the 76 UPDRS scores assigned to the LA trials considered in the experimental analysis.}
\label{testbed_UPDRS_distribution}
\end{figure}

The large-scale experimental results carried out for this work refer to a group of 24 PD patients (17 males and 7 females) with age ranging from 31 years to 79 years (with mean equal to 65.9~years and standard deviation equal to 12.3~years). The patients have been asked to perform ten repetitions of LA per leg, providing them instructions as described in Subsection~\ref{sec:LA_description}.
A total of 72 LA trials (which comprise 36 RLA and 36 LLA) have been collected.\footnote{Note that, even if only 24 patients have been considered, some patients have performed the LA task multiple times, at different times and/or for different PD conditions.}
The patients' UPDRS scores, assigned by neurologists, ranged from 0 to 3.5. To this end, note that, unlike what the MDS document indicates, non-integer ($\cdot$.5-type) scores have also been used in the case of indecision between two consecutive integer UPDRS values. In particular, the distribution of the 76 UPDRS scores assigned to the considered LA trials is shown in Fig.~\ref{testbed_UPDRS_distribution}.

\subsection{Acquisition and Optoelectronic Validation} \label{subsec:acquisition}
As anticipated in Subsection~\ref{subsec:hw}, spatial and temporal parameters, along with the kinematic of the user lower limbs, have been monitored and evaluated using the considered BSN-based inertial system and a reference optoelectronic system.
Specifically, the optoelectronic system has been used to estimate the 3D position of passive markers positioned on specific anatomical landmarks of the subject. Passive markers data were collected on all body segments (pelvis, thigh, shank, and foot bilaterally). The Davis marker-set was chosen as the protocol of choice to acquire the motion of lower limbs and trunk based on~\cite{DaOuTyGa91,FeBePaFrBeRaCrLe08}.

\begin{figure}[t]
\centering
\begin{tabular}{cc}
 \includegraphics[width=0.21\textwidth]{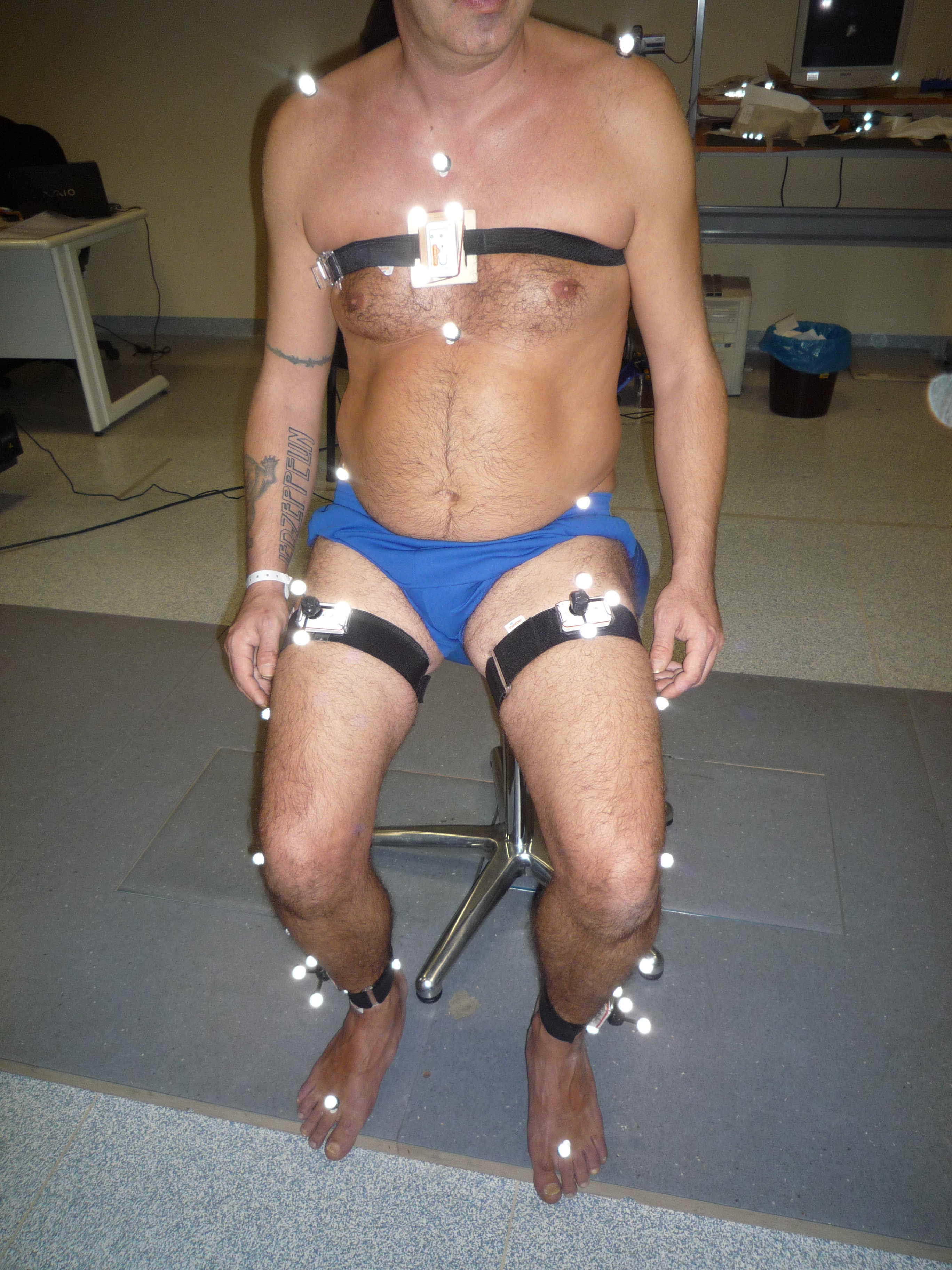} &
\includegraphics[width=0.165\textwidth]{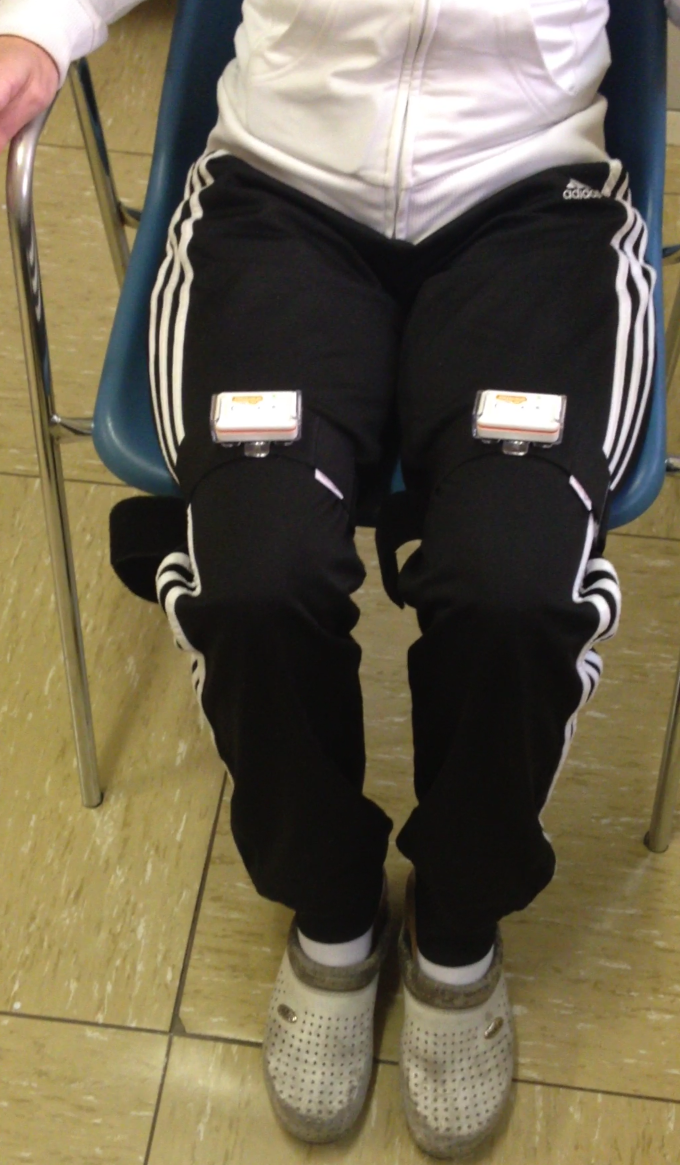} \\
(a) & (b)
\end{tabular}
\caption{Overview of the experimental testbed applied to a monitored subject: (a) configuration for single-subject
optoelectronic validation; (b) final configuration for large-scale analysis.}
\label{testbed}
\end{figure}

Concerning the inertial system, a Shimmer node (with sampling rate equal to 102.4~Hz) has been attached to each thigh of the monitored user with Velcro straps. The Shimmer devices have been placed trying to align the plane defined by the $x$ and $y$ axes of the device with the frontal plane of the user and trying to align one of the two axes with the direction of the femur. For ease of clarity, the placement of Shimmer nodes on the patient thighs is shown in Fig.~\ref{testbed}.

\begin{figure}[t]
\centering
\includegraphics[width=0.38\textwidth]{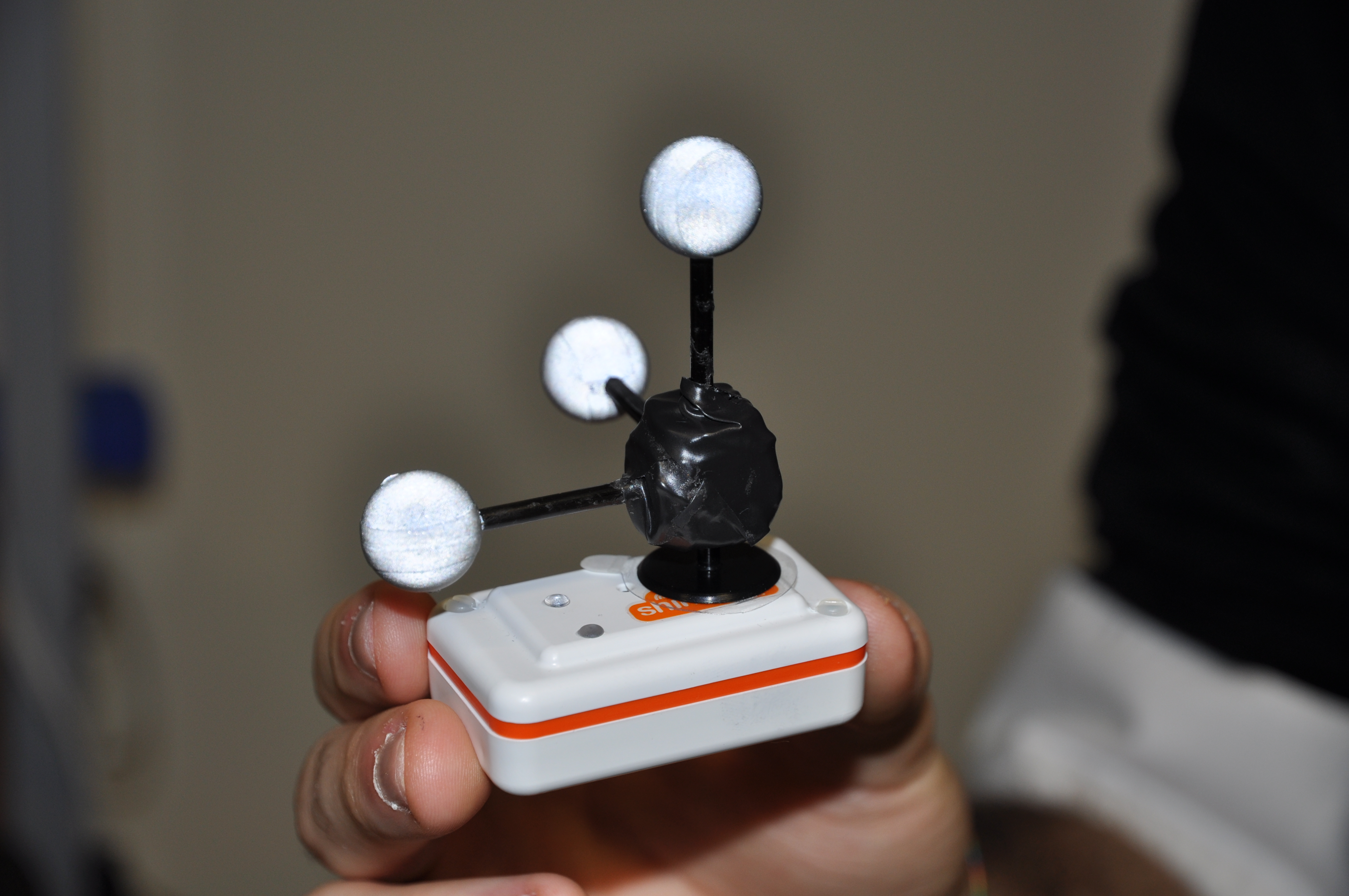}
\caption{Three orthogonally positioned reflective markers positioned over a Shimmer node.}
\label{frame}
\end{figure}

In addition to the markers specified in the Davis protocol, two groups of three markers are mounted on two frames fixed on two Shimmer devices (for a total of 6 additional markers). The three reflective markers are fixed on each Shimmer device through a frame of orthogonal rods of equal lengths (aligned, as precisely as possible, with the reference system of the Shimmer device), as shown in Fig.~\ref{frame}.
The estimation of the 3D positions of the markers of a frame (with the optoelectronic system) allows to estimate the reference orientation of the device, which is used as ground truth for the actual Shimmer orientation. We remark that, because of imperfections in the frame design (i.e., the rods could not be perfectly orthogonal or have the same lengths), a ``best-fit'' orientation is estimated~\cite{Ba00b}. Furthermore, due to a possible misalignment between the Shimmer reference system and the frame, a calibration step is performed once (at the beginning of each exercise) in order to estimate the fixed rotation between the two reference systems. This rotation is then applied to align the following measurements.

Finally, the inertial and optoelectronic systems (which are already independently synchronized) are jointly synchronized by computing the angular velocity of every optical frame and comparing it with the angular velocity measured by the corresponding Shimmer. The estimated time shift value is determined as the one which maximizes the correlation of the two signals.

\section{Leg Agility Characterization}   \label{sec:la_chara}
\subsection{Equivalence between Heel Elevation and
Thigh Inclination: Optoelectronic Validation} \label{subsec:optoelec}
According to the guidelines of the MDS, the LA task of the UPDRS should be evaluated by observing specific significant variables. As an example, the amplitude of the heel elevation and the speed of each repetition should be monitored, specifically focusing on their variations along the duration of the exercise. Furthermore, hesitations, interruptions, and freezing of the movement should also be evaluated.
As shown in Table~\ref{tab:mappingUPDRS}, general rules can be easily constructed in order to define an unambiguous mapping between observed variables and UPDRS scores.

These variables can be quite easily extracted from optical data just observing the estimated 3D positions of each marker placed on the subjects' heels (one per heel) and, in particular, its ``vertical'' component, denoted as $z_{\rm H}$ (dimension: [m]), which then indicates its elevation. To this end, a segmentation of the LA data, manually performed upon frame-by-frame observation of the videos recorded for each session, is necessary. Hesitations, interruptions, and freezing of the movement are more difficult to define but they can generally be associated with sudden variations, fluctuations, or pauses in $z_{\rm H}$ and in the linear ``vertical'' velocity $v_{\rm H}\triangleq {\rm d}z_{\rm H}/{\rm d}t$ (dimension: [m/s]).


First, we show that, in order to analyze the LA task, it is sufficient to consider the Shimmer nodes positioned over the thighs. In order to do this, we verify that the analysis of thighs' kinematic (measured through the inertial system) is actually equivalent to that of the heels' kinematic (measured through the optoelectronic system). 

To this end, the 3D orientation of a Shimmer device is estimated through an orientation filter, based on a gradient descent algorithm, which properly weighs the measurements of the three sensors of the Shimmer (i.e., accelerometer, gyroscope, and magnetometer)~\cite{MaHaVa11}. 
The inclination $\theta$ (dimension: [deg]) of the thigh is then computed as the angle between the Shimmer axis (parallel to the femur direction) and the world vertical axis, reduced by 90~deg---this is expedient to measure 0~degrees when the subject is sitting. Moreover, the angular velocity of the thighs, directly measured through the Shimmer's gyroscope, is considered. In particular, we define as $\omega$ (dimension: [deg/s]) the component of the angular velocity measured around the Shimmer axis perpendicular to the femur direction and lying in the frontal plane of the user.

At this point, $z_{\rm H}$ is compared to $\theta$ and $v_{\rm H}$ is compared to $\omega$. For both subjects~A and B introduced in Subsection~\ref{subsubsec:single_sub}, the correlation between $z_{\rm H}$ and $\theta$ is between 0.98 and 0.99 and the correlation between $v_{\rm H}$ and $\omega$ is between 0.93 and 0.98, showing then a strong correlation between heels' optical data and thighs' inertial data---no graph is shown for lack of space. Therefore, this motivates the use, in the following analysis, of the signals extracted from the Shimmer nodes of the thighs, i.e., $\theta$ and $\omega$.

\begin{figure}[t]
\centering
\includegraphics[width=0.37\textwidth]{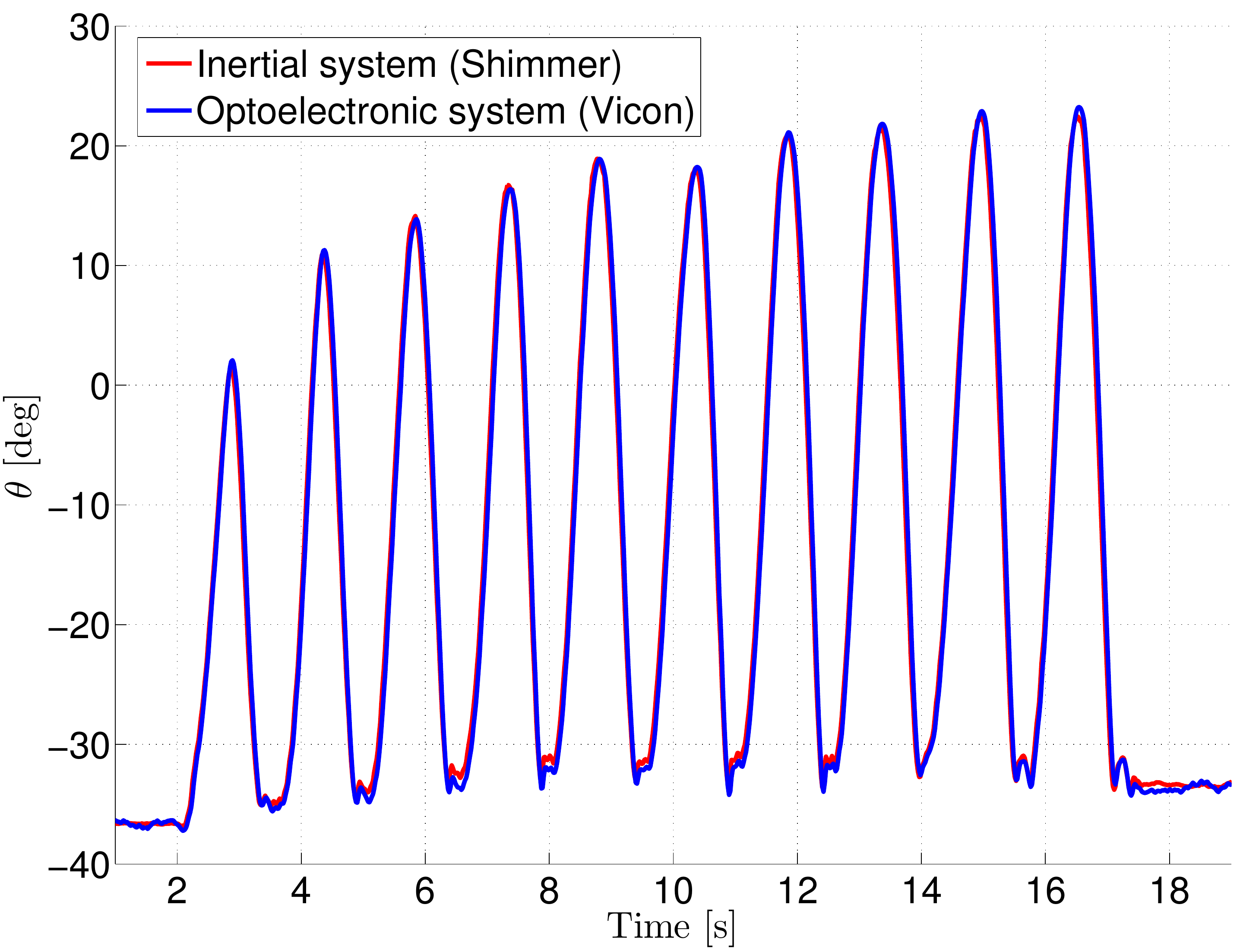}
\caption{Direct comparison between the inclinations estimated through optoelectronic (Vicon) and inertial (Shimmer) systems.}
\label{inclination}
\end{figure}

In order to highlight the accuracy in the estimation of the thigh inclination $\theta$ provided by the orientation filter acting on Shimmer data, in Fig.~\ref{inclination} a direct comparison of the inclination estimated through the optoelectronic system (by means of the orthogonal frame of markers shown in Fig.~\ref{frame}) and that estimated through the inertial sensor is shown.
It is easy to see that, even if the subject movement presents a high dynamicity, the error between the inclination measured with the two systems is almost negligible. Therefore, the analysis conducted in the following will be based on inertial data.
%

\subsection{Leg Agility Features in the Time Domain} \label{sec:time_LA_features}

Upon frame-by-frame observation of the videos recorded for each session (execution of the LA task by a patient), the inclination signals have been manually segmented in order to extract information about single repetitions within the LA task. This segmentation allows to define three time labels, denoted as $t_{\rm S}(r)$, $t_{\rm E}(r)$, and $t_{\rm P}(r)$, associated, respectively, with the start, the end, and the epoch of maximal thigh inclination of the $r$-th LA repetition ($r \in \{1,2,\dots,10\}$).

\begin{figure}[t]
\centering
\includegraphics[width=0.4\textwidth]{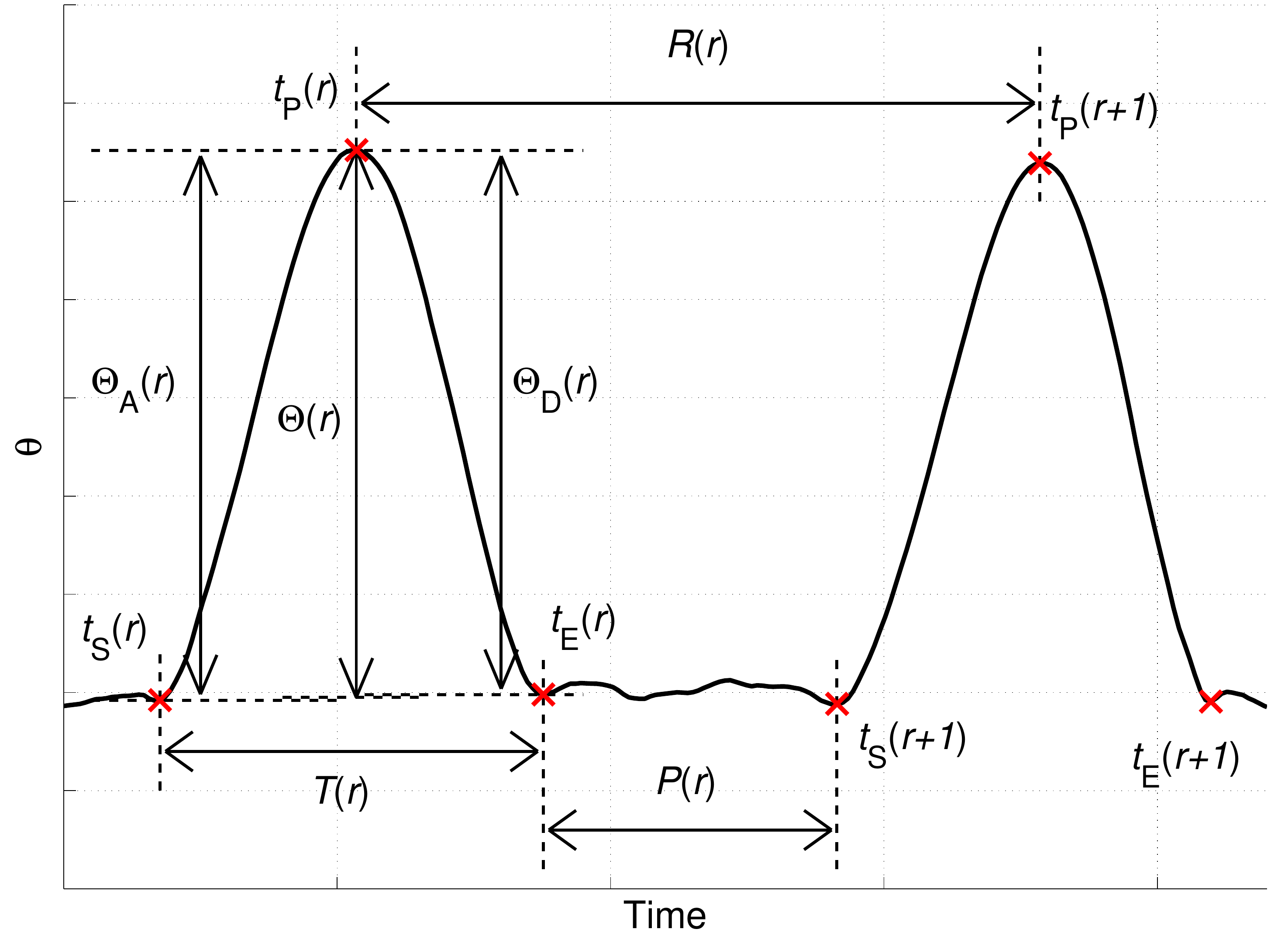}
\caption{Graphically intuitive representation of some of the involved kinematic variables. Two illustrative consecutive repetitions of LA are shown together with the corresponding segmentation events (red crosses).}
\label{LA_features}
\end{figure}

Starting from the previous labels and the inclination signal, for each repetition $r$ ($r \in \{1,2,\dots,10\}$) of the LA (either RLA or LLA) the following features, relative to the time domain, can be straightforwardly computed.
\begin{itemize}
 \item The angular amplitude $\Theta(r)$ (dimension: [deg]):
\begin{equation*}
\Theta(r) \triangleq \frac{\Theta_{\rm A}(r)+\Theta_{\rm D}(r)}{2}
\end{equation*}
where
\begin{eqnarray}
\Theta_{\rm A}(r)& \triangleq &\theta(t_{\rm P}(r))-\theta(t_{\rm S}(r))
\\
\Theta_{\rm D}(r)& \triangleq &\theta(t_{\rm P}(r))-\theta(t_{\rm E}(r)).
\end{eqnarray}
 \item The angular speed of execution $\Omega(r)$ (dimension: [deg/s]):
\begin{equation*}
\Omega(r)\triangleq\frac{\Theta_{\rm A}(r)+\Theta_{\rm D}(r)}{T(r)}=\frac{\Theta_{\rm A}(r)+\Theta_{\rm D}(r)}{t_{\rm E}(r)-t_{\rm S}(r)}.
\end{equation*}
\end{itemize}
Furthermore, for each pair of consecutive repetitions, say $r$ and $r+1$ ($r \in \{1,\dots,9\}$), of the LA (either RLA or LLA) the following features can be computed.
\begin{itemize}
 \item The $r$-th pause of execution $P(r)$ (dimension: [s]):
\begin{equation*}
P(r) \triangleq t_{\rm S}(r+1)-t_{\rm E}(r).
\end{equation*}
 \item The $r$-th regularity of execution $R(r)$ (dimension: [s]):
\begin{equation*}
R(r) \triangleq t_{\rm P}(r+1)-t_{\rm P}(r).
\end{equation*}
\end{itemize}
For ease of clarity, in Fig.~\ref{LA_features} some of the just introduced kinematic variables are shown for two illustrative consecutive repetitions of LA, say $r$ and $r+1$.
Finally, if we jointly consider the RLA and the LLA repetitions of the same patient, for each repetition $r$ ($r \in \{1,2,\dots,10\}$) of the LA the following features can also be  computed.
\begin{itemize}
 \item The relative difference of the angular amplitude between LLA and RLA $D_{\rm \Theta,LR}(r)$ (adimensional: [\%]):
\begin{equation*}
D_{\rm \Theta,LR}(r)\triangleq \frac{\Theta_{\rm LLA}(r)-\Theta_{\rm RLA}(r)}{\Theta_{\rm RLA}(r)}\cdot100.
\end{equation*}
 \item The relative difference of the angular speed between LLA and RLA $D_{\rm \Omega,LR}(r)$ (adimensional: [\%]):
\begin{equation*}
D_{\rm \Omega,LR}(r)\triangleq \frac{\Omega_{\rm LLA}(r)-\Omega_{\rm RLA}(r)}{\Omega_{\rm RLA}(r)}\cdot100.
\end{equation*}
\end{itemize}

Since a single LA repetition may lead to misleading indications about the execution of the whole LA task, the following average features (obtained by averaging, over consecutive repetitions, the features introduced above) are considered in the following experimental analysis.\footnote{From a notation viewpoint, the symbols used for average features
are the same of those of repetition-based features, but the dependence from $r$ disappears.}
\begin{eqnarray*}
\Theta \triangleq \frac{\sum_{r=1}^{10} \Theta(r)}{10} && 
\Omega\triangleq\frac{\sum_{r=1}^{10} \Omega(r)}{10}\\
P\triangleq\frac{\sum_{r=1}^{9} P(r)}{9} &&
R\triangleq\frac{\sum_{r=1}^{9} R(r)}{9}\\
D_{\rm \Theta,LR}\triangleq\frac{\sum_{r=1}^{10} D_{\rm \Theta,LR}(r)}{10} &&
D_{\rm \Omega,LR}\triangleq\frac{\sum_{r=1}^{10} D_{\rm \Omega,LR}(r)}{10}.
\end{eqnarray*}
In addition, the following standard deviations (with the same dimensions) of some of the previous features are also considered:
\begin{eqnarray*}
\Theta_{\rm SD} \triangleq \sqrt{\frac{\sum_{r=1}^{10} (\Theta(r)-\Theta)^2}{9}} && 
\Omega_{\rm SD} \triangleq \sqrt{\frac{\sum_{r=1}^{10} (\Omega(r)-\Omega)^2}{9}}\\
P_{\rm SD} \triangleq \sqrt{\frac{\sum_{r=1}^{9} (P(r)-P)^2}{8}} &&
R_{\rm SD} \triangleq \sqrt{\frac{\sum_{r=1}^{9} (R(r)-R)^2}{8}}\\
\end{eqnarray*}

We also introduce the repetition frequency $F$ (dimension: [Hz]), defined as follows:
\begin{equation*}
 F\triangleq\frac{10}{t_{\rm E}(10)-t_{\rm S}(1)}.
\end{equation*}
Note that, for each LA trial, two values of $\Theta$, $\Omega$, $P$, $R$, and $F$ are computed (namely, one for the RLA and one for the LLA), whereas only one value of $D_{\rm \Theta,LR}$ and $D_{\rm \Omega,LR}$ is computed.\footnote{In the large-scale analysis considered in Subsection~\ref{subsec:res-large_scale}, 72 values of $\Theta$, $\Omega$, $P$, $R$, and $F$ will be available, whereas 36 values of $D_{\rm \Theta,LR}$ and $D_{\rm \Omega,LR}$ will be available.
Finally, a UPDRS score $u \in \{0,0.5,1,1.5,2,2.5,3,3.5,4\}$ will be assigned to each of the 72 LA trials.}

\subsection{Leg Agility Features in the Frequency Domain} \label{sec:frequency_LA_features}
While the features considered in Subsection~\ref{sec:time_LA_features} belong to the time domain, we now focus on the characterization of the LA task by extracting relevant information on the UPDRS value in the frequency domain.

For the following frequency analysis, the signals $\theta(t)$ (i.e., the inclination of the thighs) and $\omega(t)$ (i.e., the angular velocity of the thighs), introduced in Subsection~\ref{subsec:optoelec}, are considered. Specifically, the latter is directly measured through the Shimmer gyroscope and is defined as the component of the angular velocity measured around the Shimmer axis perpendicular to the femur direction and lying in the frontal plane of the user.
The considered signals $\theta$ and $\omega$ are properly segmented in order to start at the initial instant of the first LA repetition and to end at the final instant of the last LA repetition.

The spectra (namely, the Discrete Fourier Transforms, DFTs) of $\theta$ and $\omega$ (properly centered on their means) have been computed using a Fast Fourier Transform (FFT) algorithm for every leg of every patient.
More formally, denoting $x\in\{\theta,\omega\}$, the $h$-th component of the spectrum $X_{\rm FFT}$ of the signal $x$ is computed as follows:
\begin{equation}
X_{{\rm FFT},h}=\sum_{n=0}^{N-1} x_n e^{-jh \frac{2\pi}{N} n} \qquad  h=0,\dots,N-1
\end{equation}
where $N$ is the length of $x$ (and, consequently, the length of $X_{{\rm FFT},h}$)
The amplitude spectrum (which we will just consider in the following) is then easily computed by taking the absolute value of $X_{{\rm FFT}}$ divided by $N$ as follows
\begin{equation}
X=\frac{\left|X_{\rm FFT}\right|}{N}.
\end{equation}

\begin{figure*}[t]
\centering
\begin{tabular}{cc}
\includegraphics[width=0.4\textwidth]{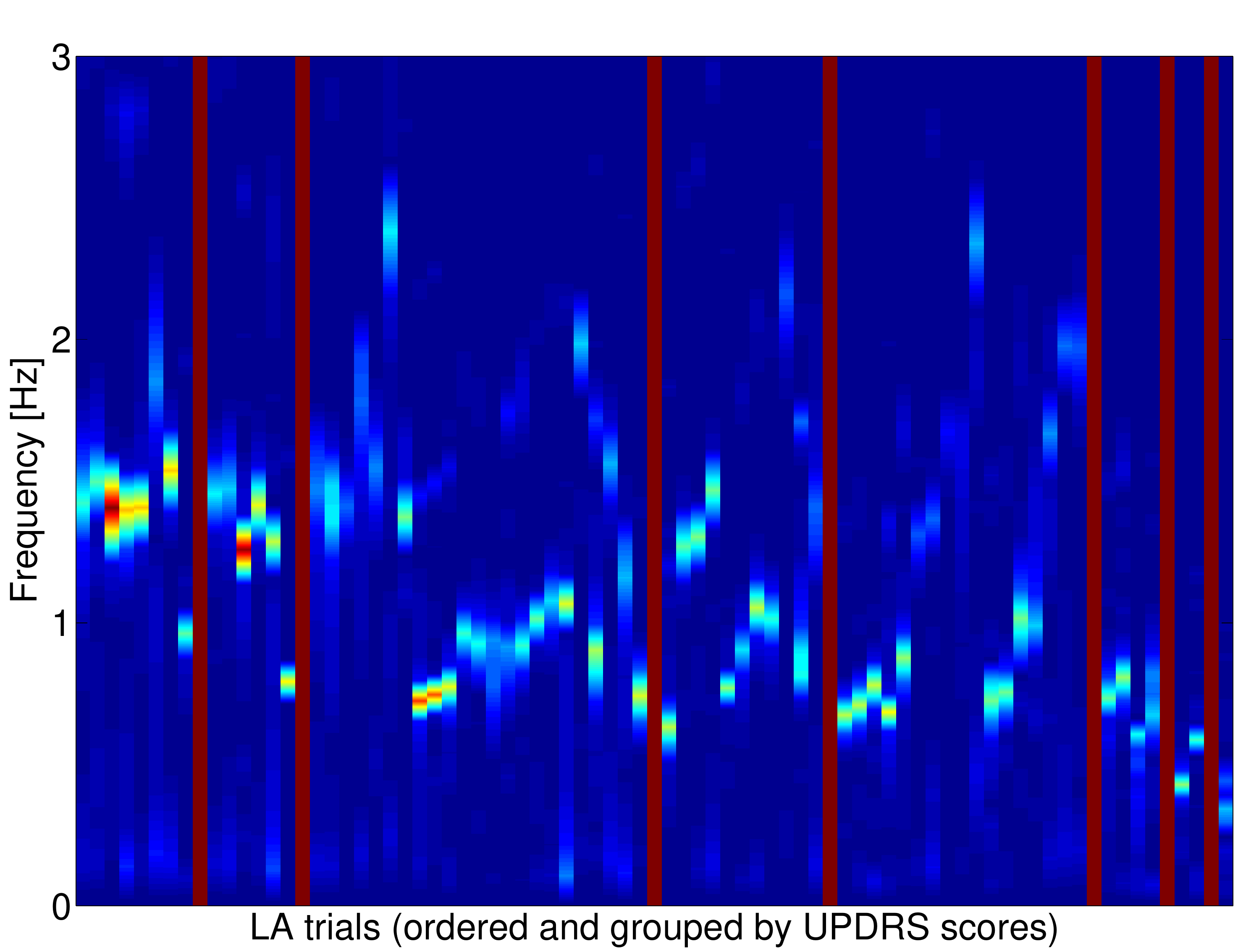} & 
\includegraphics[width=0.4\textwidth]{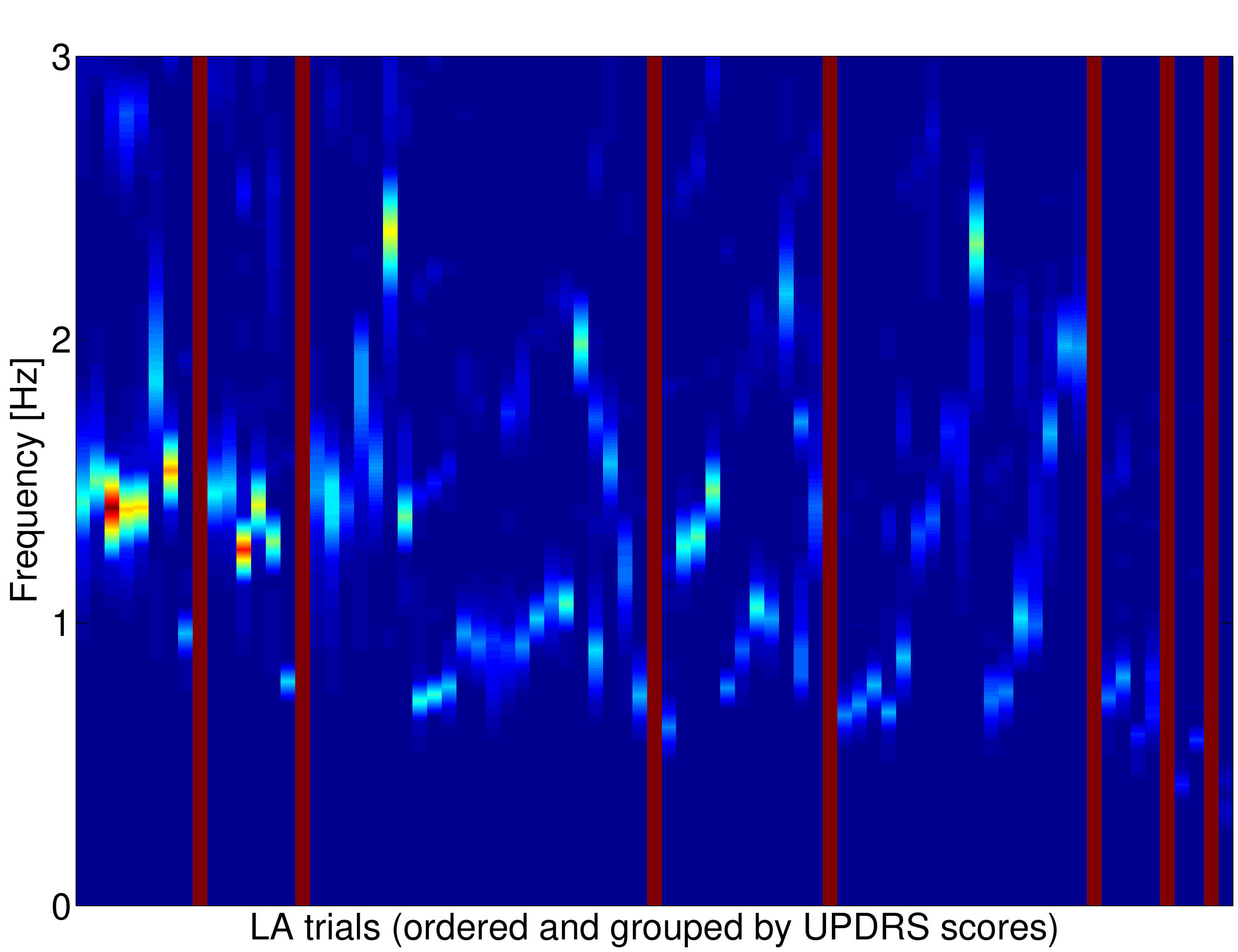} \\
(a)-$\theta$ & (b)-$\omega$ \\
\includegraphics[width=0.4\textwidth]{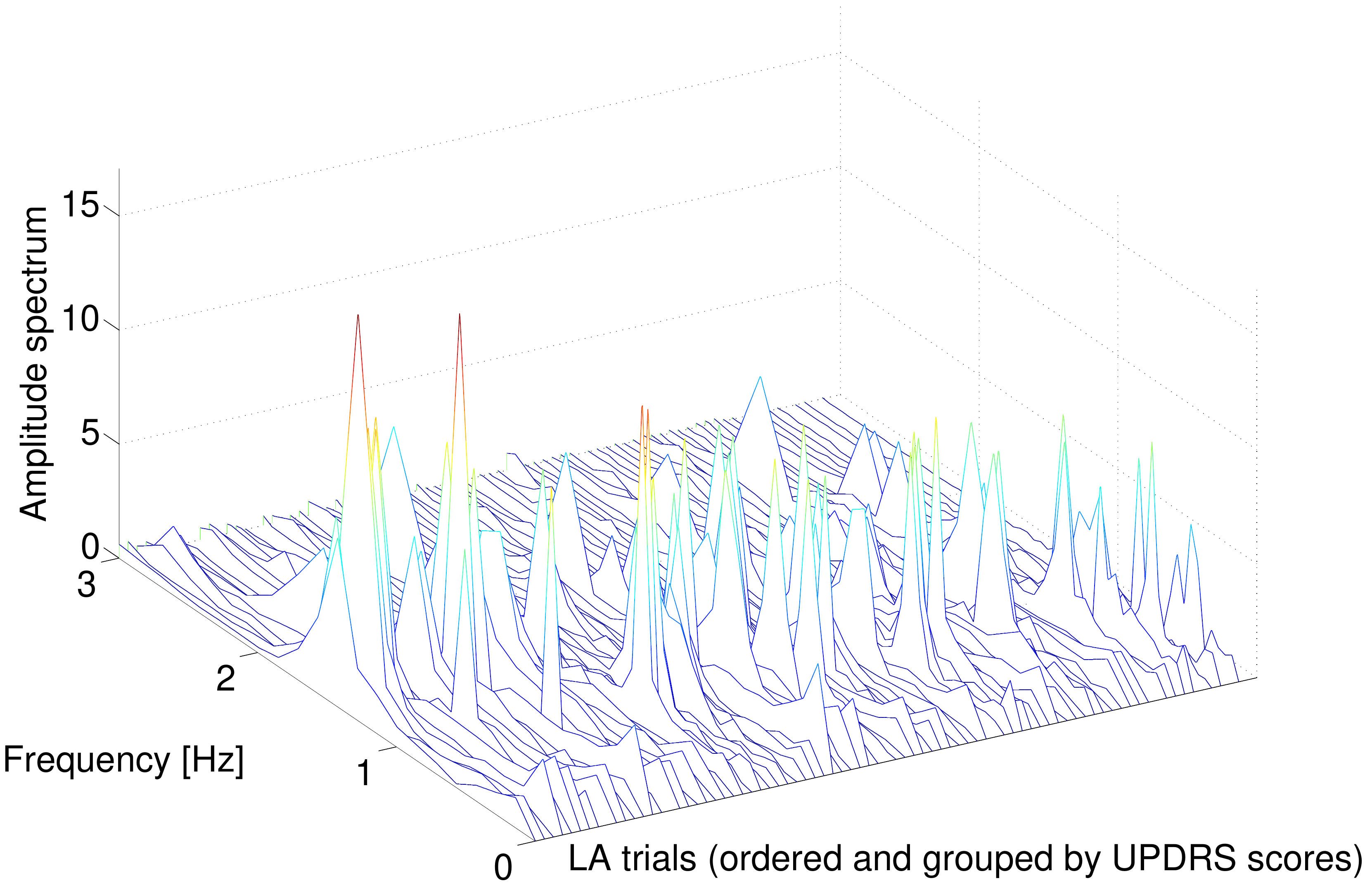} &
\includegraphics[width=0.4\textwidth]{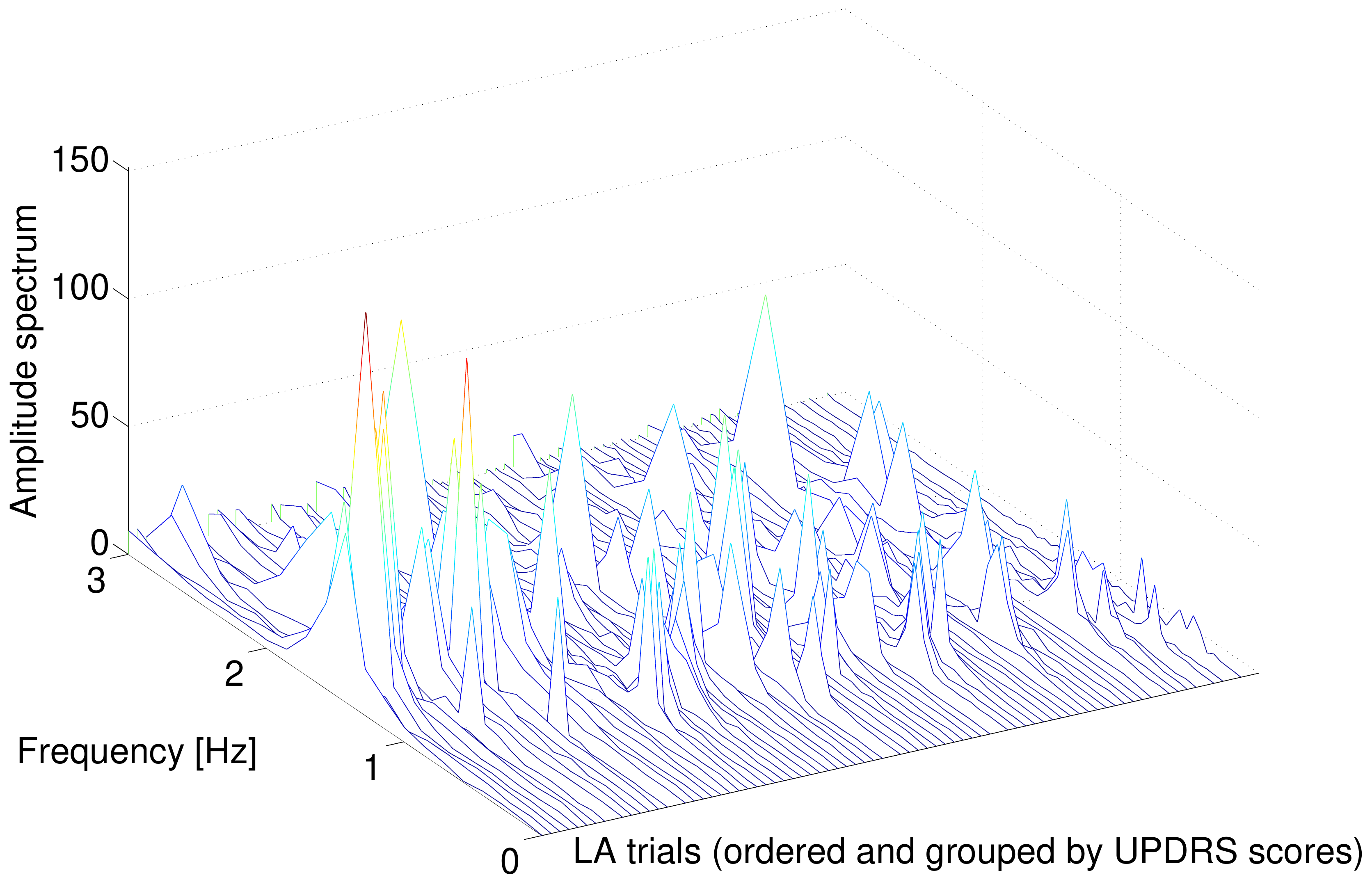} \\
(c)-$\theta$ & (d)-$\omega$
 \end{tabular}
\caption{One-sided amplitude spectra of $\theta$ (i.e., inclination of the thigh) and $\omega$ (i.e., angular velocity of the thigh) for different LA trials: (a)-(b) two-dimensional and (c)-(d) three-dimensional representations. Magnitudes of the spectra are mapped to a color that ranges from blue (lowest values) to red (highest values). The LA trials are ordered and grouped by UPDRS scores (from 0 to 3.5). Groups of spectra corresponding to different UPDRS scores are separated either (a)-(b)~using a vertical red line or (c)-(d)~skipping a line in the graphs.}
\label{incl_spectra}
\end{figure*}

The obtained spectra have been then grouped and ordered according to the UPDRS score of the corresponding patient.
All the computed spectra of $\theta$ and $\omega$ are shown, using a one-sided representation, in Fig.~\ref{incl_spectra}.
For these spectra, both two-dimensional (subfigures~(a)-(b)) and three-dimensional (subfigures~(c)-(d)) representations are provided.
It may be observed that the spectra amplitudes for both $\theta$ and $\omega$ generally decrease moving from low UPDRS scores to higher ones.
To this end, in our analysis we will take into account a new feature, i.e., the spectrum power.
More formally, for the spectrum $X$, its power $P_{X}$ is computed as follows:
\begin{equation}
P_X= \frac{1}{N} \sum_{h=0}^{N-1} \left(X_h\right)^2.
\end{equation}
In the following, for ease of understanding when $X$ is the spectrum of either $\theta$ or $\omega$, the used notation will be $X_{\theta}$ or $X_{\omega}$, respectively.

\section{Results and Discussion} \label{sec:results}
\subsection{Single-subject Analysis}
\begin{figure}[t]
\centering
\begin{tabular}{c}
\includegraphics[width=0.38\textwidth]{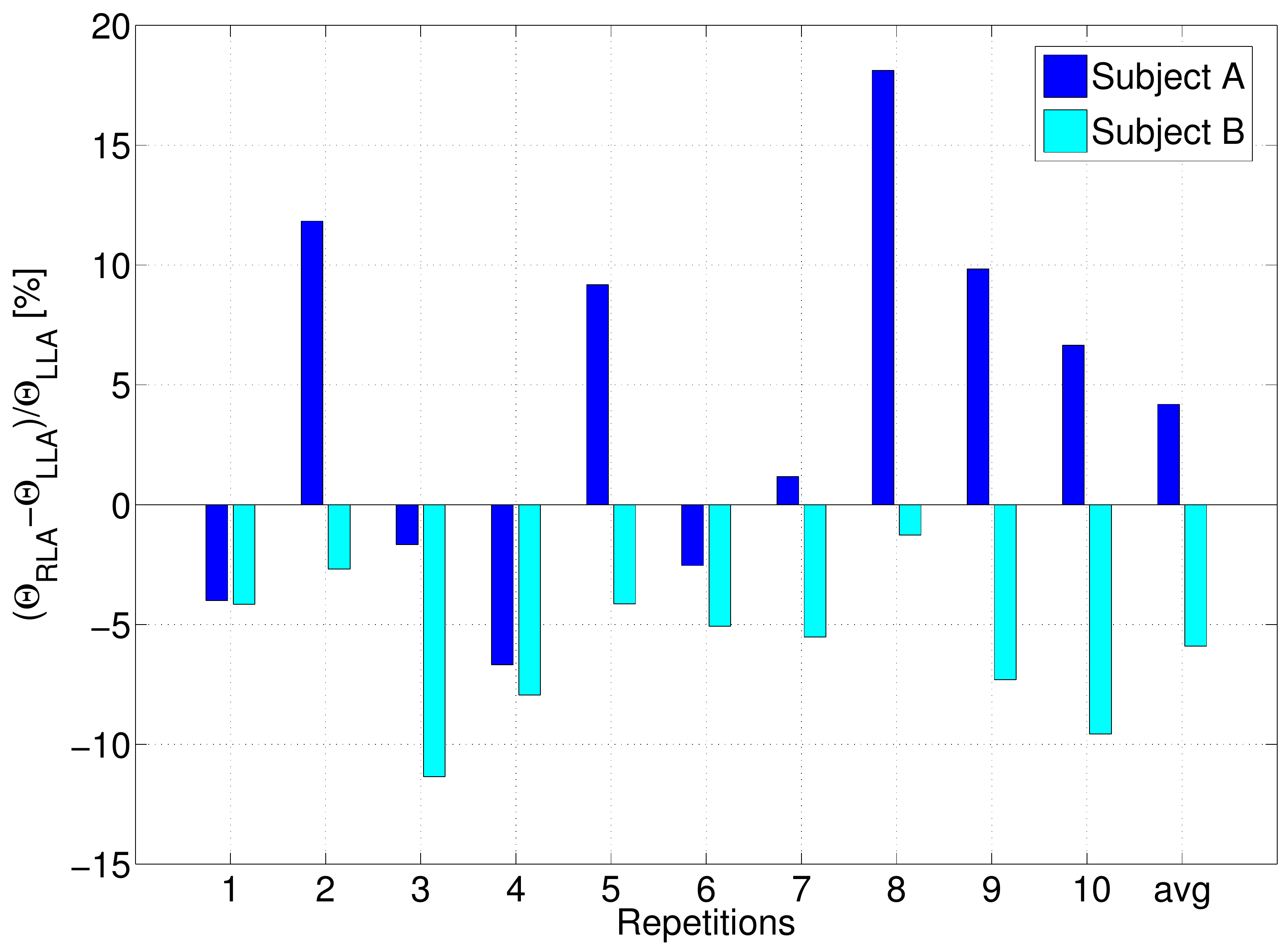}\\
(a) \\
 \includegraphics[width=0.38\textwidth]{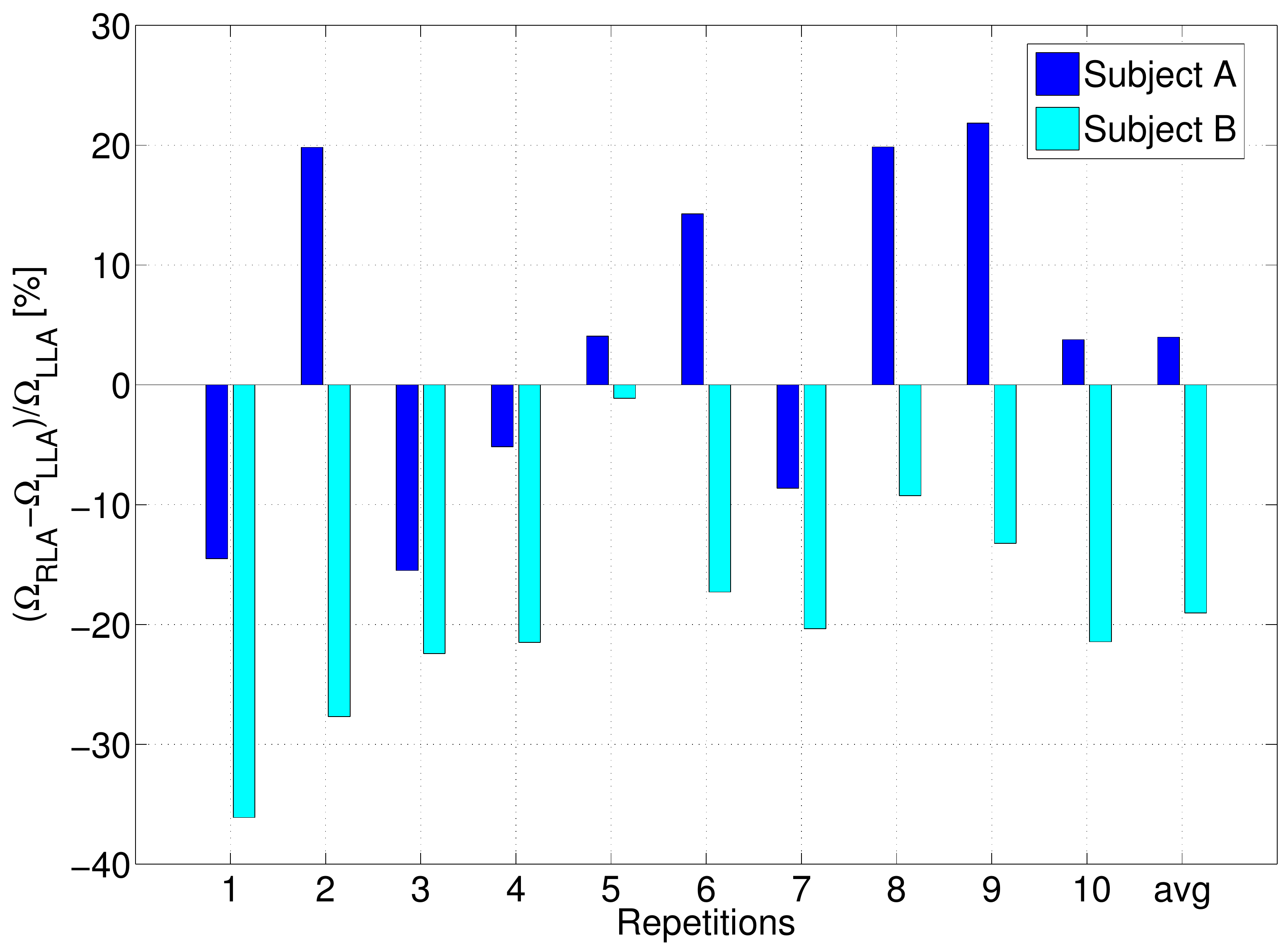} \\
(b)
\end{tabular}
\caption{Comparison, over ten repetitions, between Subject~A (healthy) and Subject~B (PD), in terms of: (a) relative angular amplitude difference (percentage) between the two legs; (b) relative angular speed difference (percentage) between the two legs.}
\label{amplitude_speed}
\end{figure}

First, we analyze the amplitude and the speed of each repetition, indicated in the MDS's UPDRS document as the main variables to observe in the LA task.  In  Fig.~\ref{amplitude_speed}~(a), the difference in angular amplitude, expressed in percentage, is shown for each repetition (from 1 to 10), comparing directly Subject~A and Subject B. 
It can be noticed that Subject~A does not present a biased difference between RLA and LLA, whereas Subject B's RLA angular amplitude is always lower than that of LLA. On average, it can be observed that the relative difference between RLA and LLA is around 4\% for Subject~A and around 6\% for Subject~B. Similarly, the difference (expressed in percentage) between the RLA and LLA angular speeds, shown in Fig,~\ref{amplitude_speed}~(b), reveals that Subject~B's RLA angular speed is 19\% lower than his LLA angular speed. On the contrary, this is generally not true for Subject A, for which the relative difference is, on average, around 4\%.

According to the results in Fig.~\ref{amplitude_speed}, it is worth to highlight that the difference observed between the RLA and the LLA of a specific subject, even if not specifically referenced in the MDS's document as a significant variable, can instead represent a clear evidence of a non-0 UPDRS score. Therefore, a possible extension of the UPDRS can be already envisioned just by introducing this new variable in the LA analysis. 

\begin{figure}[t]
\centering
\begin{tabular}{c}
\includegraphics[width=0.38\textwidth]{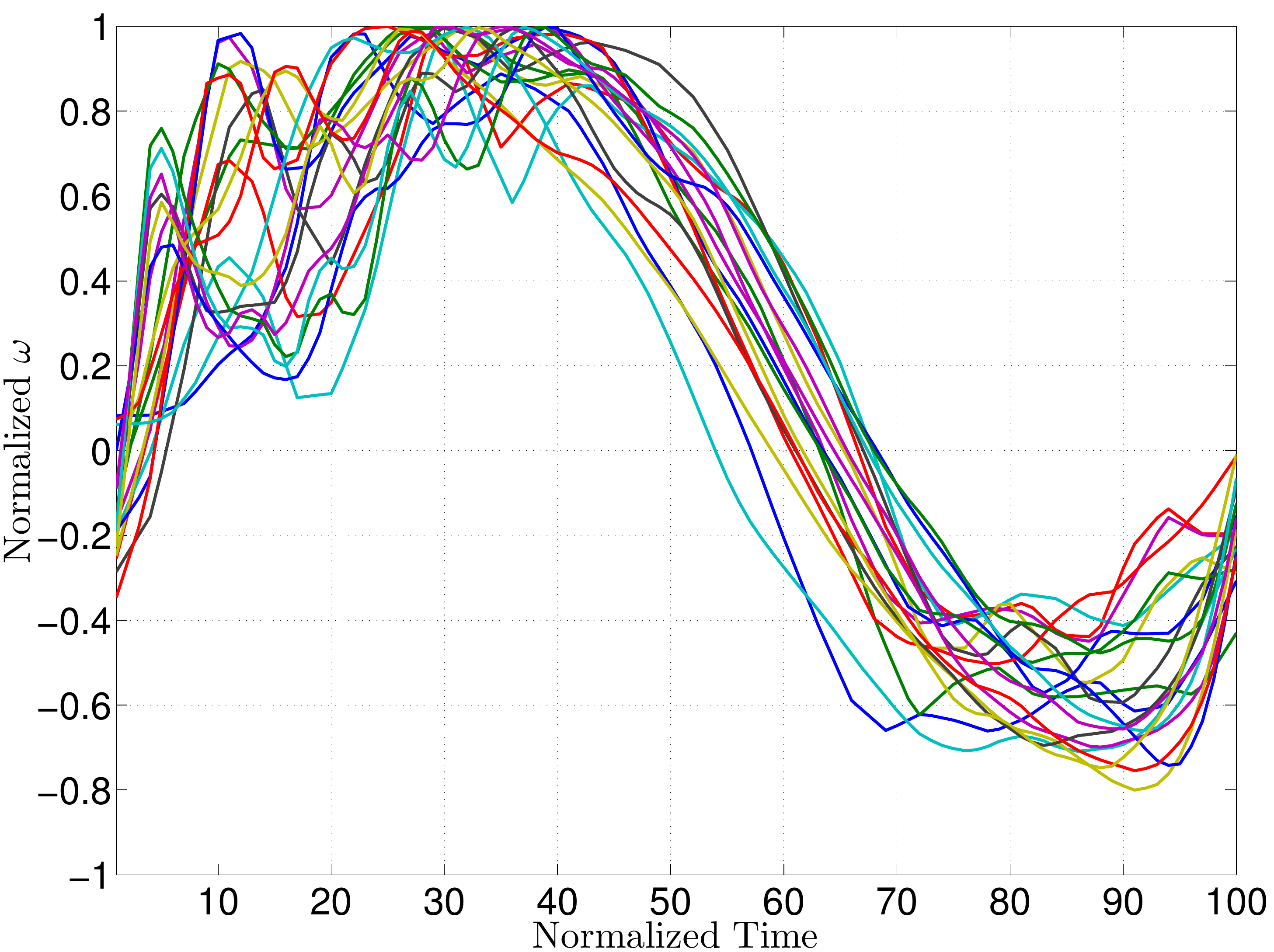} \\
(a) \\
\includegraphics[width=0.38\textwidth]{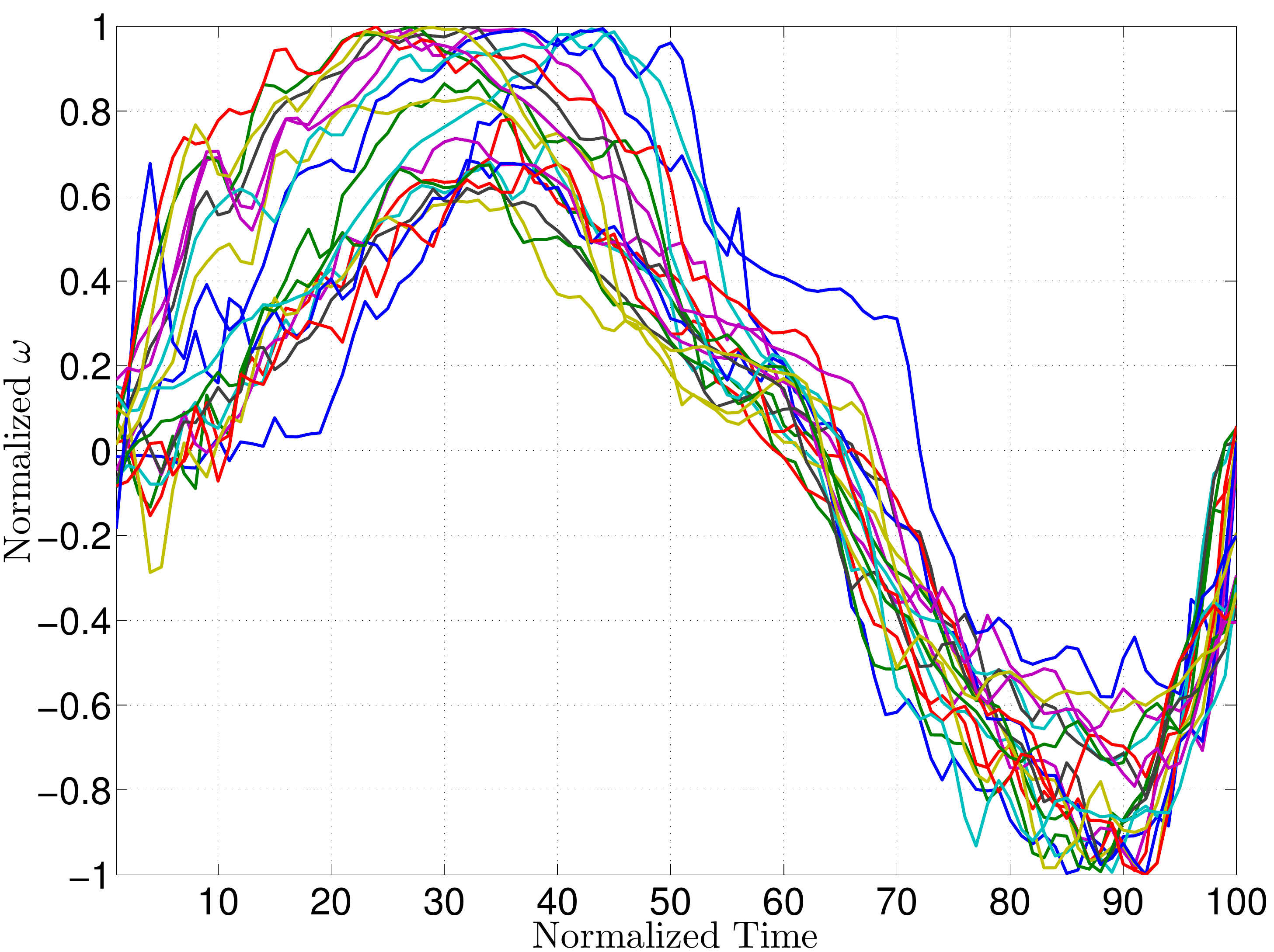} \\
(b)
\end{tabular}
\caption{Juxtaposed normalized (between $\pm 1$) angular velocities $\omega$ of ten repetitions, normalized over time (between 1 and 100) for (a)~Subject A and (b)~Subject B.}
\label{omega_rep}
\end{figure}

In order to better investigate and characterize the LA repetitions, a qualitative analysis has been carried out by investigating the angular velocity $\omega$ along an entire repetition. In particular, in Fig.~\ref{omega_rep}, the segmented portions of $\omega$, corresponding to each LA repetition, have been normalized in time and value (so that time goes from 1 to 100 and $-1\leq \omega \leq +1$) and overlapped for (a)~Subject A and (b)~Subject B, respectively.
Even if both subjects show some macro-hesitations (specifically when raising up the heels), it is easy to see that Subject~B's angular velocity presents several micro-hesitations, not easily observable, with a naked eye, in recorded videos. This can be then another useful kinematic aspect to be taken into account in order to better investigate the LA task and predict the presence of PD. 

\subsection{Large-scale Analysis} \label{subsec:res-large_scale}
As anticipated in Subsection~\ref{subsec:paper_cont}, the aim of the following analysis is to devise an approach to automatically assign a UPDRS score to a specific LA task. To this end, it is crucial to determine if there exists a relationship between the UPDRS score assigned by neurologists to an LA task and the values of the kinematic variables introduced in Section~\ref{sec:la_chara}. In particular, our goal is to determine if there exist UPDRS ``trajectories,'' in the kinematic (multi-dimensional) feature space, which clearly allow to detect the correct UPDRS value.

\begin{figure}[t]
\centering
\includegraphics[width=0.4\textwidth]{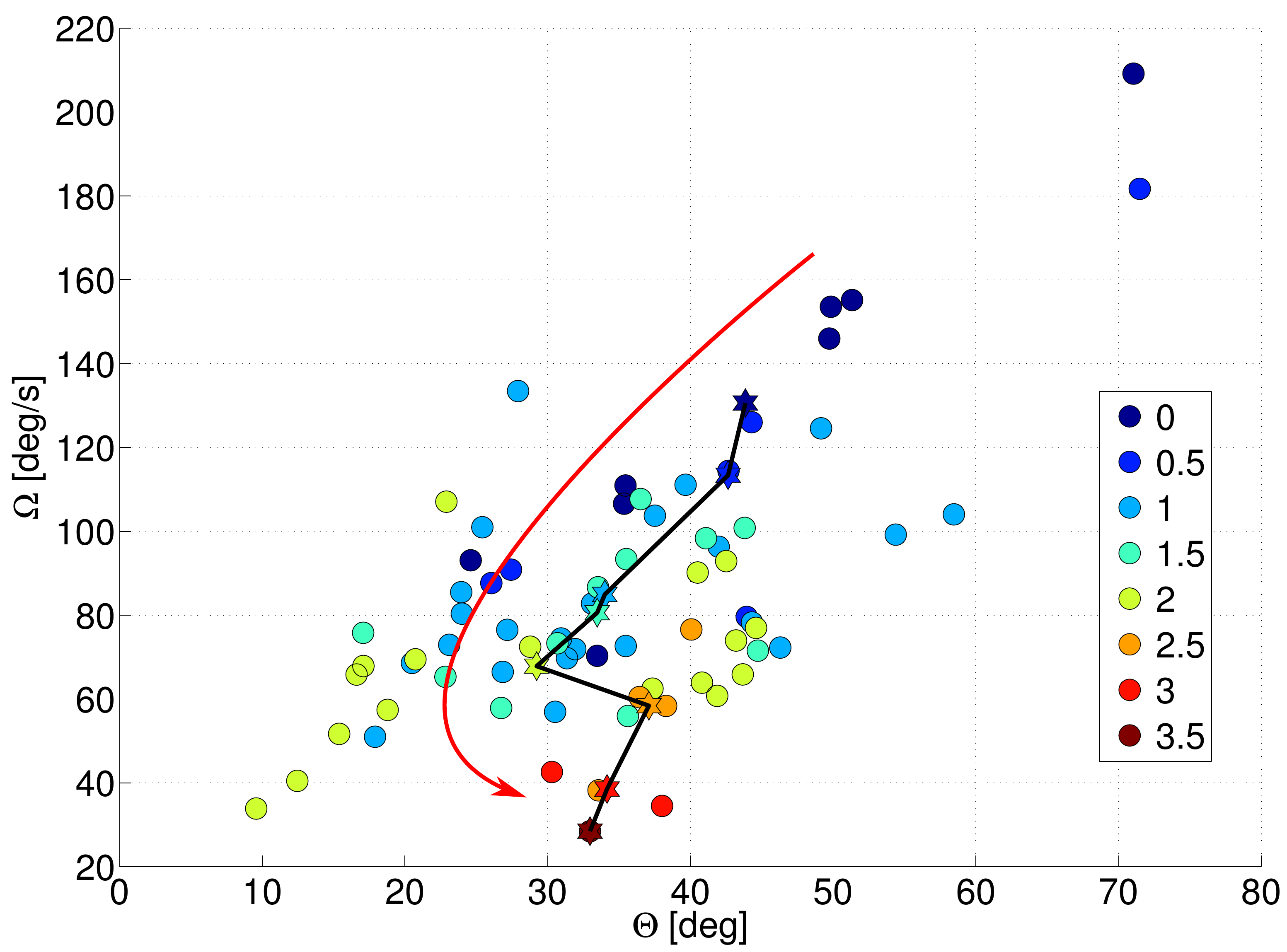}\\
\caption{Representation of  average amplitude-speed pairs $\{(\Theta_i,\Omega_i)\}_{i=1}^{72}$ on the same plane. Pairs are colored according to the corresponding UPDRS scores. Centroids of each cluster of pairs (drawn as filled stars), corresponding to the same UPDRS score, are shown and are linked in UPDRS-wise order from 0 to 3.5. For ease of clarity, a representative red arrow shows a smoothed version of this trajectory.}
\label{amplSpeed_pauseReg}
\end{figure}

In Fig.~\ref{amplSpeed_pauseReg}, all the average amplitude-speed pairs $\{(\Theta_i,\Omega_i)\}_{i=1}^{72}$ are represented on the same plane.
Each pair is assigned a color corresponding to the related UPDRS score. Furthermore, the centroids (denoted by the filled stars) of each cluster of pairs, corresponding to the same UPDRS score, are also shown. Finally, the black line links the centroids from 0 to 3.5 (in addition, a representative red arrow shows a smoothed version of this trajectory). It can be observed that uniform clusters of pairs with same UPDRS scores are highly distinguishable for very low and very high UPDRS scores. Furthermore, even if clusters of pairs corresponding to different UPDRS scores tend to overlap, a clear trend is visible showing that pairs mainly move from the upper right corner (i.e., the pairs with UPDRS 0) to the bottom part (i.e., the pair with UPDRS 3.5) of the plane. This also supports the intuition that amplitude and speed of motion are mostly correlated.

\begin{figure}[t]
\centering
\includegraphics[width=0.4\textwidth]{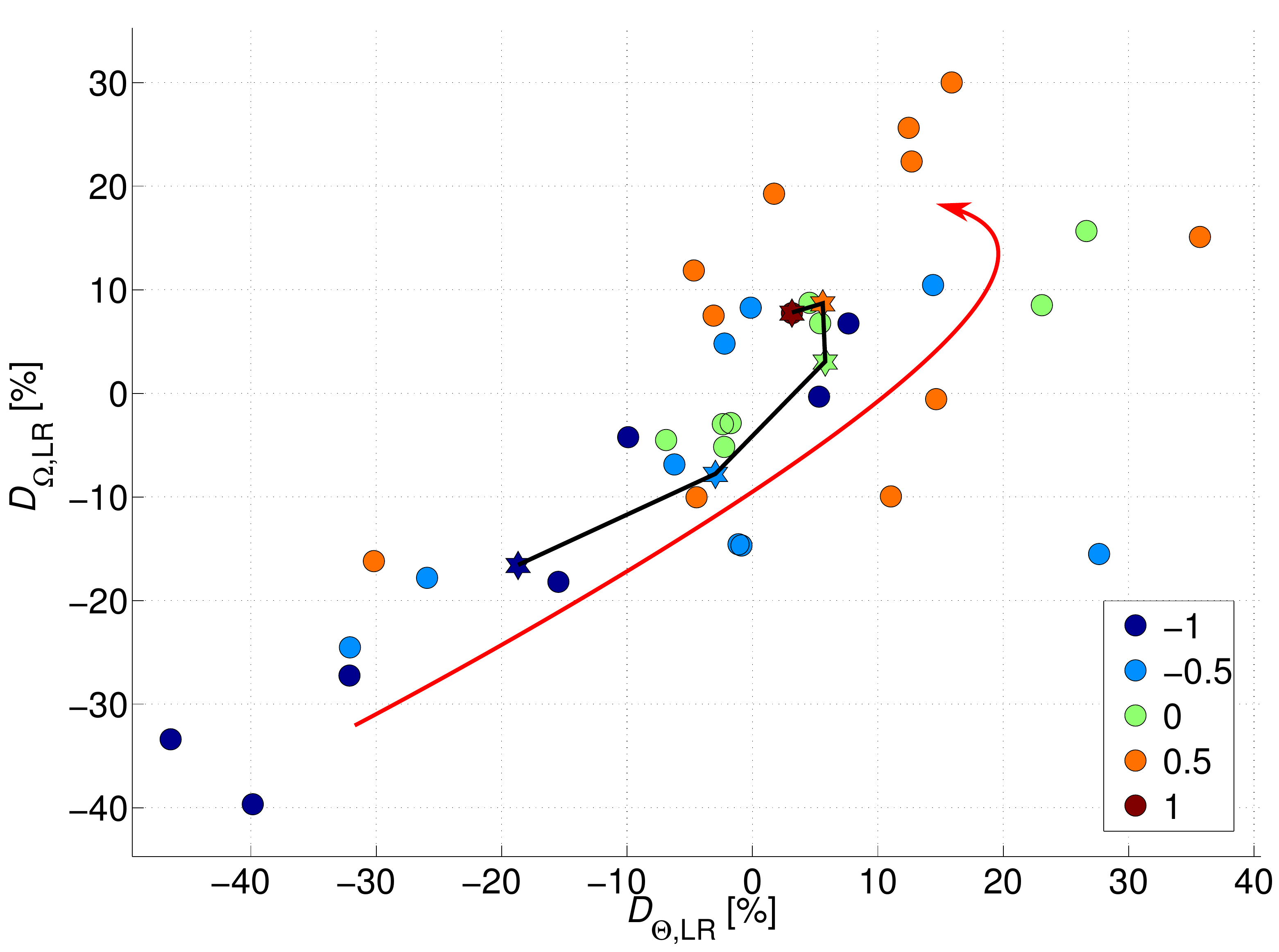} 
\caption{Representation of average relative differences of amplitude-speed pairs between LLA and RLA $\{(D_{\rm \Theta,LR,i},D_{\rm \Omega,LR,i})\}_{i=1}^{36}$ on the same plane. Pairs are colored according to the corresponding difference of UPDRS scores between the left and right legs. Centroids of each cluster of pairs (drawn as filled stars), corresponding to the same UPDRS score difference, are shown and are linked in ascending order from -1 to 1. For ease of clarity, a representative red arrow shows a smoothed version of this trajectory.}
\label{diff}
\end{figure}

Focusing now on the joint analysis of the RLA and the LLA, in Fig.~\ref{diff} the average relative differences of amplitude-speed pairs between LLA and RLA $\{(D_{\rm \Theta,LR,i},D_{\rm \Omega,LR,i})\}_{i=1}^{36}$ are drawn on the same plane. 
As before, a clear trend is visible in the graph. Furthermore, it can be observed that low UPDRS scores are associated with pairs placed around (0,0). On the other hand,
the highest differences in UPDRS scores correspond to pairs which are distant from the (0,0) point, i.e., to LA tasks where significant differences between RLA and LLA emerge.

So far, it has been shown that there is a relationship between the UPDRS scores and the measured values of the kinematic variables. However, for ease of visualization, we have considered just pairs of variables. In order to overcome this limitation and consider all kinematic variables simultaneously, in~\cite{GiFeCoCiAzAlMa14} a Principal Component Analysis (PCA) can be considered, in order to reduce the dimensionality of the measured data while still retaining most of the variance of the original data~\cite{DuHaSt00}. In~\cite{GiFeCoCiAzAlMa14}, PCA is used to map the input five-dimensional points $\{(\Theta_i,\Omega_i,P_i,R_i,F_i)\}_{i=1}^{72}$ into a lower-dimensional space.\footnote{Note that, since different kinematic variable values have different magnitudes and ranges, before applying PCA, the original data are first centered at their means (which are set equal to 0) and rescaled to have a unit standard deviation.} According to the results of the considered PCA, 87.8\% of the original variance is retained by just using the first two components, whereas its 92.7\% is retained by adding the third component. This shows, as expected, the presence of redundancy in the measured data. Such redundancy can be significantly reduced by just considering the original data projected onto two-dimensional or three-dimensional spaces, as determined by the PCA.

In order to evaluate the performance of an automatic detection system able to associate a measured set of kinematic values to a specific UPDRS score, three methods have been considered: Nearest Centroid Classifier (NCC), $k$ Nearest Neighbors ($k$NN), and Support Vector Machine (SVM)~\cite{DuHaSt00}. In a few words: the NCC method classifies a new (unknown) point according to the same label of the nearest centroid (in terms of Euclidean distance); the $k$NN method classifies the new point according to the labels of the $k$ nearest points (still in terms of Euclidean distance) through a majority rule; and the SVM method classifies the new point according to decision regions (associated with the UPDRS classes) that are constructed in order to maximize separation between different classes.
In order not to bias the performance of the classifiers, a leave-one-out cross-validation method is considered. This means that each point of the original dataset is used, in turn, as the new (unknown) point and the remaining points are used as a training dataset. The performance is then evaluated by averaging together the single observed performance results.
Finally, concerning the $k$NN method, even if different values of $k$ have been considered and evaluated, in the following we will just consider the case with $k=4$ (which, heuristically, appears to optimize the system performance).
We also define the absolute UPDRS error $e$ as follows:
\begin{equation*}
e \triangleq |\widehat{u}-u|
\end{equation*}
where $u$ is the actual UPDRS score and $\widehat{u}$ is the estimated one (using NCC, $k$NN, or SVM).

\begin{figure}[t]
\centering
\includegraphics[width=0.4\textwidth]{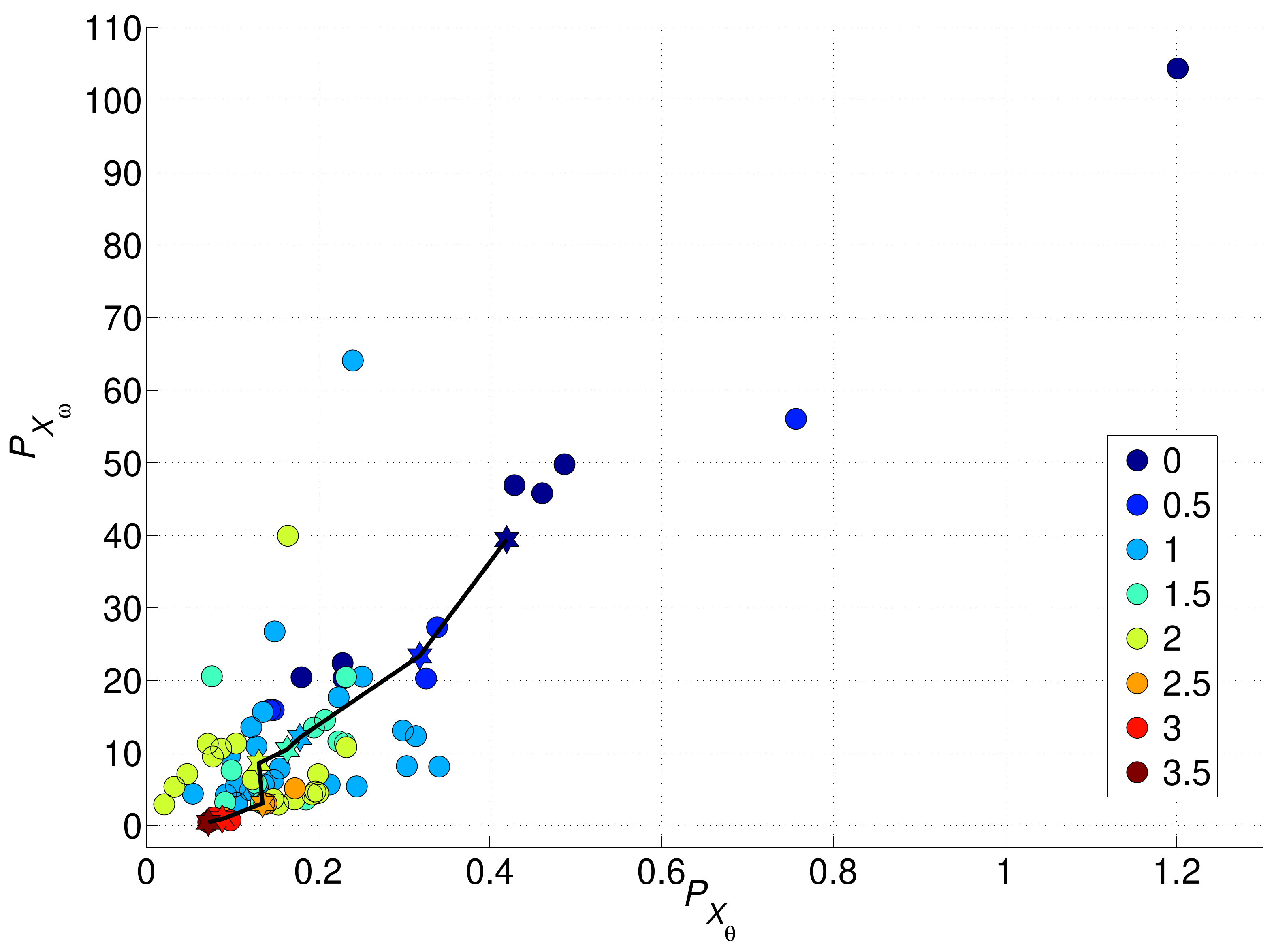}
\caption{Representation of pairs $\{(P_{X_\theta,i},P_{X_\omega,i})\}_{i=1}^{72}$ on the same plane. Pairs are colored according to the corresponding UPDRS scores. Centroids of each cluster of pairs (drawn as filled stars), corresponding to the same UPDRS score, are shown and are linked in UPDRS-wise order from 0 to 3.5.}
\label{inclPow_angVelPow}
\end{figure}

In Fig.~\ref{inclPow_angVelPow}, all the pairs $\{(P_{X_\theta,i},P_{X_\omega,i})\}_{i=1}^{72}$, which take into account the powers of the spectra of the frequency features introduced in Subsection~\ref{sec:frequency_LA_features}, are represented on the same plane.
As for the previous figures, each pair is assigned a color corresponding to the related UPDRS score. Furthermore, the centroids (denoted by the filled stars) of each cluster of pairs, corresponding to the same UPDRS score, are also shown. Finally, the black line links the centroids from 0 to 3.5. Note that, even if clusters of pairs corresponding to different UPDRS scores tend to overlap, a clear trend is visible showing that pairs mainly move from the upper right corner (i.e., the pairs with UPDRS 0) to the bottom part (i.e., the pair with UPDRS 3.5) of the plane.

\begin{figure}[t]
 \centering
 \begin{tabular}{c}
\includegraphics[width=0.45\textwidth]{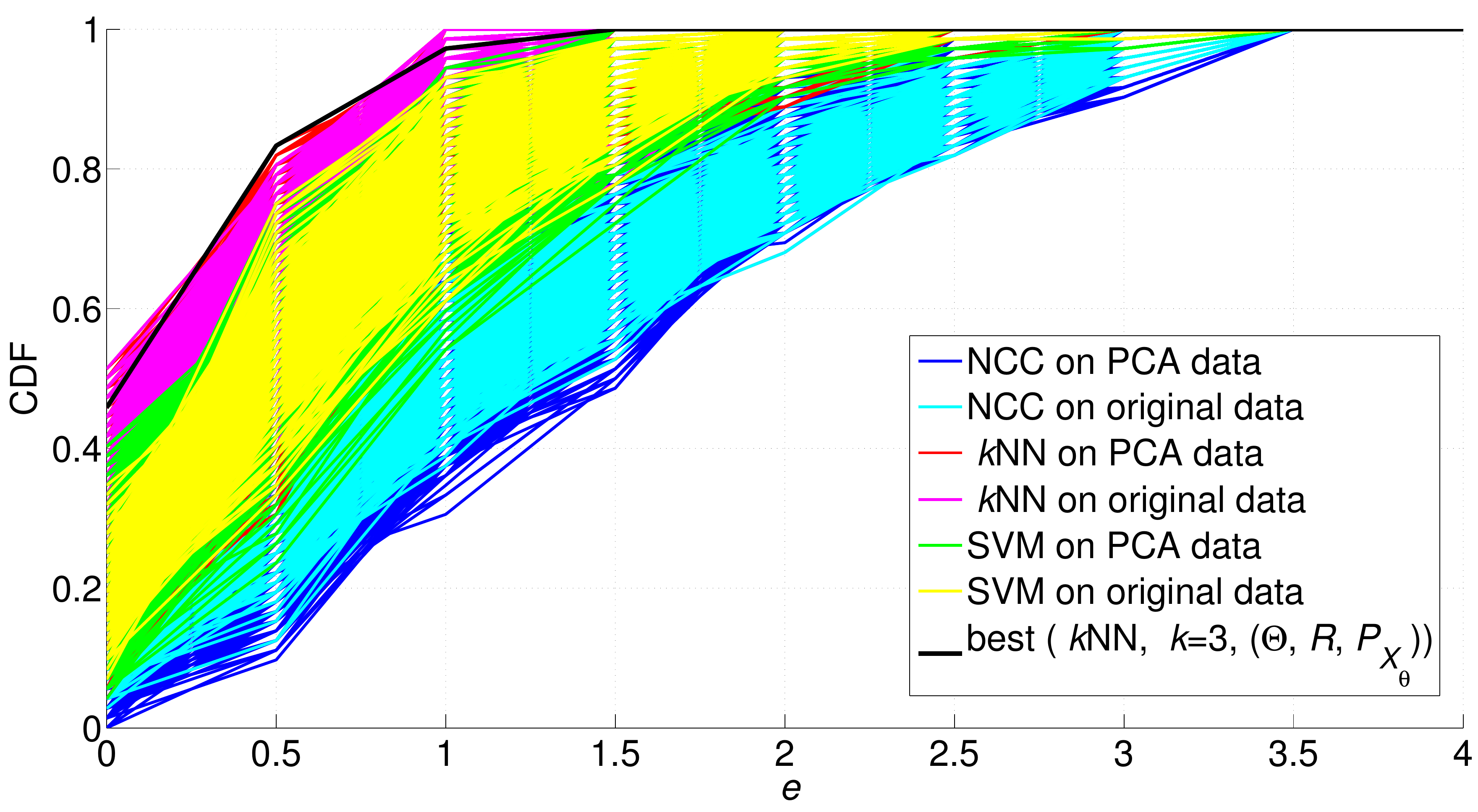} \\
(a) \\
\includegraphics[width=0.45\textwidth]{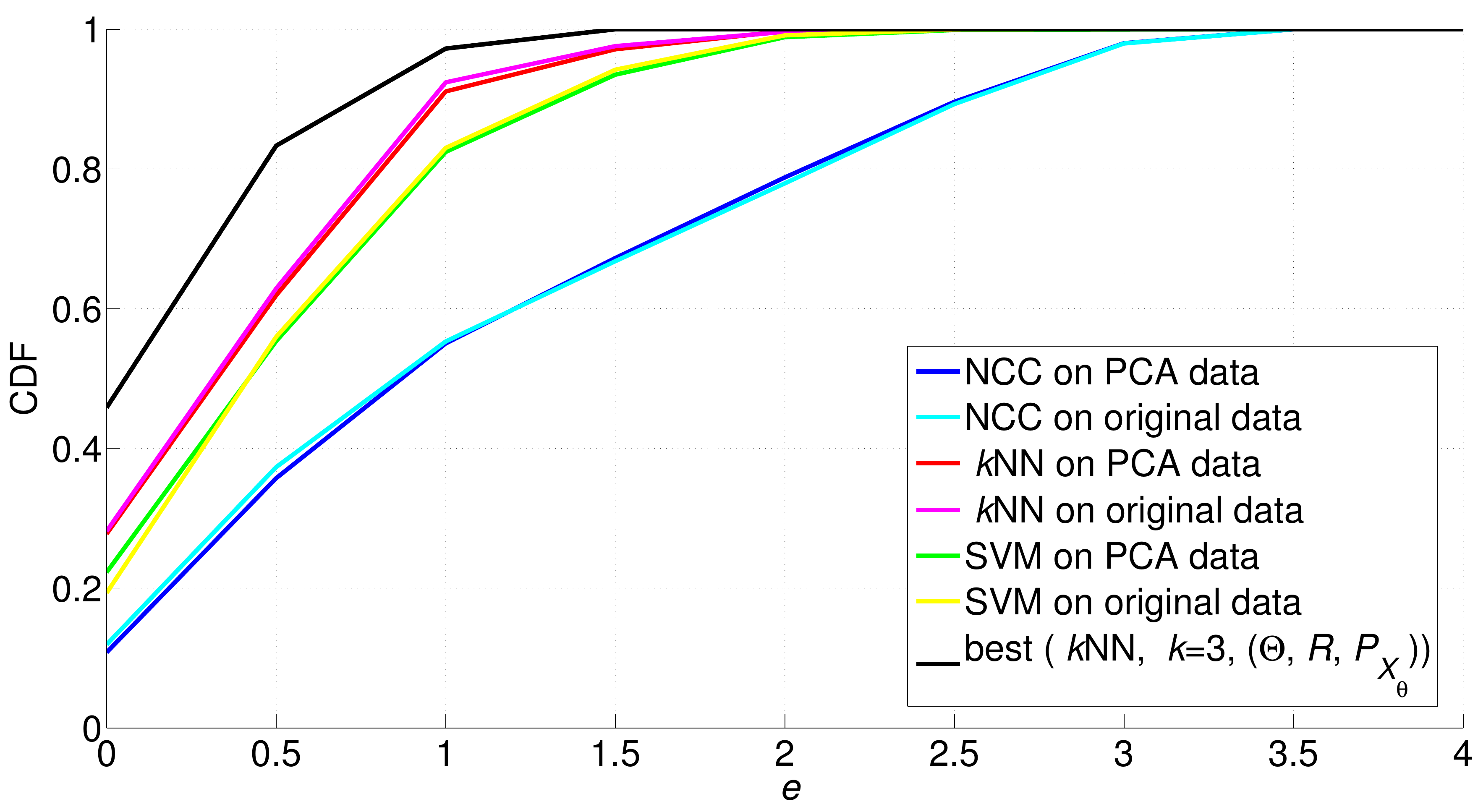} \\
(b)
\end{tabular}
 \caption{CDFs of the absolute UPDRS error $e$ using NCC, $k$NN, and SVM. The previous classifiers are used on both the original data and the PCA-projected data. In (a), the CDFs for all possible combinations of parameters and features are shown, whereas, in (b), the average CDFs for every classifier are shown. The black solid line is the CDF of the best case (i.e., $k$NN with $k=3$, using ($\Theta$,$R$,$P_{X_{\theta}}$) as features).}
 \label{error_distrib}
 \end{figure}

\begin{figure}[t]
 \centering
  \includegraphics[width=0.4\textwidth]{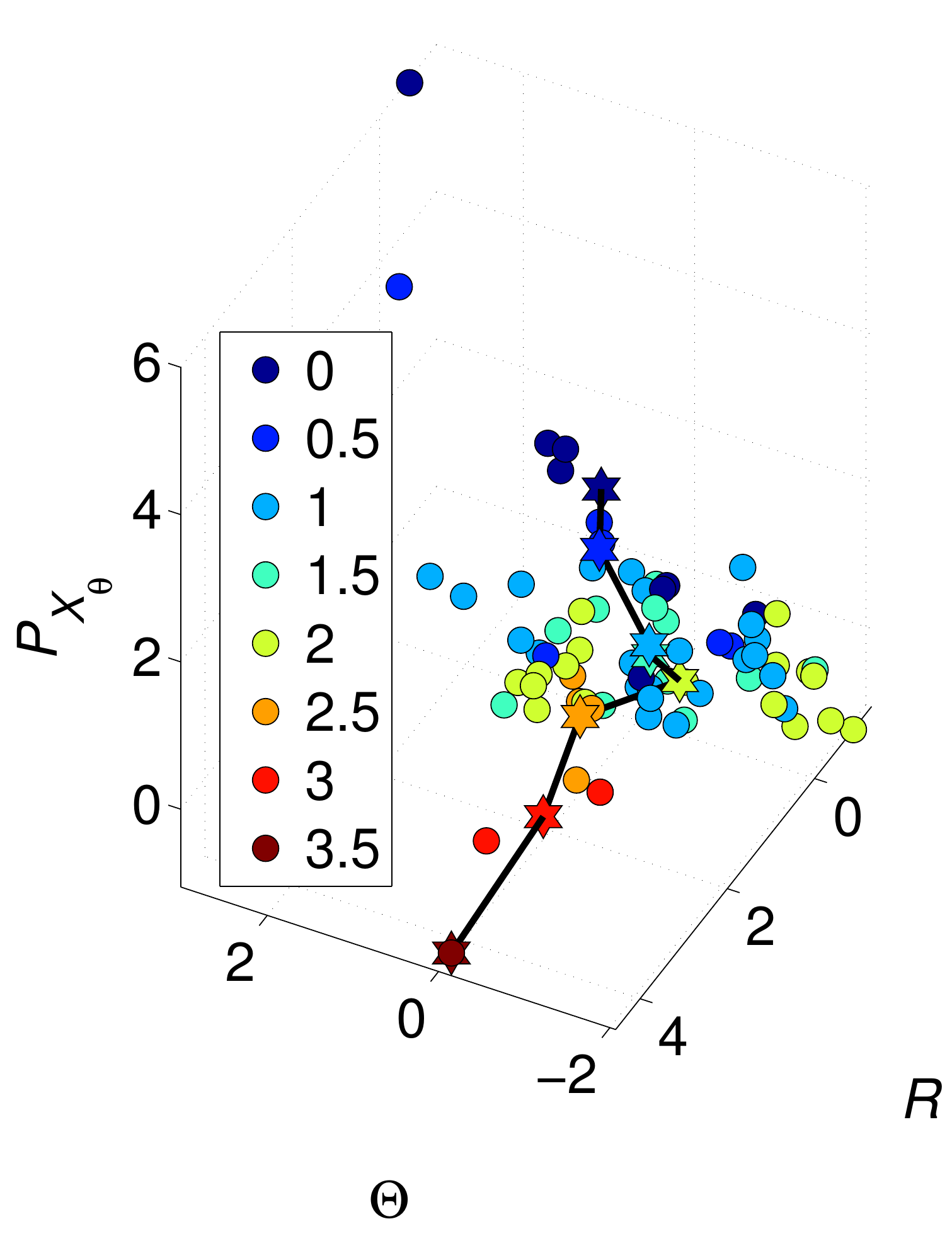}
 \caption{Representation of 3D points $\{\Theta_i,R_i,P_{X_\theta,i}\}_{i=1}^{72}$ on the same space. 3D points are colored according to the corresponding UPDRS scores. Centroids of each cluster of points (drawn as filled stars), corresponding to the same UPDRS score, are shown and are linked in UPDRS-wise order from 0 to 3.5.}
 \label{bestComb}
 \end{figure}

In order to investigate the best performance achievable with the proposed system and with the considered features, an exhaustive performance analysis has been carried out, by testing: the system performance for all possible combinations of features; possible values of $k$ (when the $k$NN method, which will turn out to be the best, is used); and the number of considered principal components (when PCA data are used, instead of original data).
In particular:
\begin{itemize}
 \item the following 11 features, among those presented earlier, are selected: $\Theta$, $\Omega$, $P$, $R$, $\Theta_{\rm SD}$, $\Omega_{\rm SD}$, $P_{\rm SD}$, $R_{\rm SD}$, $F$, $P_{X_{\omega}}$, and $P_{X_{\theta}}$;
 \item when using the $k$NN method, the following values of $k$ are considered: $1,2,\dots,10$;
 \item up to 11 principal components (as the number of features) are used when considering PCA data.
\end{itemize}
The three presented classifiers (namely, NCC, $k$NN, and SVM) have been run  on both original data and ``PCA-projected'' data.
In Fig.~\ref{error_distrib}, a direct comparison of the Cumulative Distribution Functions (CDFs) of $e$ with NCC, $k$NN, and SVM is carried out. 
In particular, in Fig.~\ref{error_distrib}~(a) the CDFs for all possible combinations of parameters and features are shown using different colors for each classifier. It can be observed that the choice of the classifier, more than the choice of the features and parameters, is crucial in order to achieve good performance. Indeed, groups of CDFs for each classifier tend to lie in the same portion of the plane even for different features and parameters. To this end, in Fig.~\ref{error_distrib}~(b) the average CDFs (over all possible combinations of features and parameters) for each classifier are also shown. It is easy to observe that best performance can be achieved using $k$NN, followed by SVM and, eventually, NCC. Note also that performing the analysis on PCA data, rather than on original data, does not significantly improve the classifiers' performance. The best combination of features and parameters, chosen as the one which maximize the Area under the Curve (AuC) of the CDFs, is resulted to be that which uses $k$NN with $k=3$ and ($\Theta$,$R$,$P_{X_{\theta}}$) as features. In the previous figures, the black solid line represents the corresponding CDF. For completeness, in Fig.~\ref{bestComb}, a trajectory, similar to the ones shown in Fig~\ref{amplSpeed_pauseReg}, Fig.~\ref{diff}, and Fig.~\ref{inclPow_angVelPow}, is drawn in the kinematic space of the best features.
As for the previous bi-dimensional figures, in this case as well a clear trajectory can be identified. This further confirms the feasibility of the design and implementation of an automatic UPDRS detection system.

\section{Conclusions}\label{sec:conc}
In this paper, we have investigated how kinematic variables, collected through a simple BSN during the LA task, can be
representative of the UPDRS value assigned by neurologists. We have first considered a ``single-subject'' perspective,
trying to highlight similarities and differences between characteristic kinematic variables of a healthy subject and
of a PD patient. Then, we have carried out a ``large-scale'' experimental investigation considering
24 PD patients. Many kinematic features, in both time and frequency domains, have been investigated, considering various classification methods (NCC, $k$NN, and SVM). The most representative
ones have been characterized considering two-dimensional (amplitude-speed and amplitude power-speed power) and three-dimensional feature spaces (amplitude-amplitude power-regularity). In order to highlight the presence of correlation between the considered kinematic variables, a PCA has been carried out. However, our results show that using PCA-processed data, rather than original data, does not significantly improve improve the classifiers' performance. The best system configuration, chosen as the one which maximize the AuC of the CDFs of the decision error, turned out to be the one based on $k$NN with $k=3$ and ($\Theta$,$R$,$P_{X_{\theta}}$) as features.  In particular, characteristic ``trajectories'' (from low to high UPDRS scores) emerge in the considered multi-dimensional kinematic spaces, which is a promising result for the design of an automated UPDRS detection system which could rely on the introduction of proper ``UPDRS decision regions'' in the considered kinematic spaces. Last, but not least, the use of a wireless BSN makes the proposed system directly integrable into IoT systems, paving the way to telerehabilitation applications and cloud computing-based processing of a large amount of heterogeneous data.


%


\section*{Acknowledgment}
This work is partially supported by the Italian Ministry of Health (RF-2009-1472190).

\ifCLASSOPTIONcaptionsoff
  \newpage
\fi

\end{document}